\documentclass[12pt]{iopart}
\usepackage{iopams}
\pdfoutput=1
\expandafter\let\csname equation*\endcsname=\relax
\expandafter\let\csname endequation*\endcsname=\relax
\usepackage{color}
\usepackage[breakwords]{truncate}
\usepackage[font={small}]{caption}
\usepackage{physics}
\usepackage{graphicx}
\usepackage{cite} 
\usepackage{subcaption}
\usepackage{etoolbox}
\usepackage{verbatim}
\usepackage[section]{placeins}
\usepackage[sectionbib]{chapterbib}
\usepackage{multirow}
\usepackage{pbox}
\usepackage{calc}
\usepackage{braket}
\usepackage[extra]{tipa}
\usepackage{csquotes}
\usepackage[breaklinks,colorlinks = true,urlcolor=blue,citecolor=blue]{hyperref}
\usepackage{calc}

\usepackage{accents}
\usepackage{romannum}
\AtBeginDocument{\pagenumbering{arabic}}

\makeatletter
\renewcommand\@appendixstar{\@@par
 \ifnumbysec 
 \@addtoreset{table}{section}
 \@addtoreset{figure}{section}\fi
 \setcounter{section}{0}
 \setcounter{subsection}{0}
 \setcounter{subsubsection}{0}
 \setcounter{equation}{0}
 \setcounter{figure}{0}
 \setcounter{table}{0}
  \def\thesection{\Alph{section}} 
 \def\theequation{\ifnumbysec
      \Alph{section}.\arabic{equation}\else
      \Alph{section}\arabic{equation}\fi}
 \def\thetable{\ifnumbysec
      \Alph{section}\arabic{table}\else
      A\arabic{table}\fi}
 \def\thefigure{\ifnumbysec
      \Alph{section}\arabic{figure}\else
      A\arabic{figure}\fi}}
\makeatother

\begin{document}
\newcommand{\ben}[1]{\textcolor{blue}{\textbf{#1}}}
\title[Current fluctuations in an interacting active lattice gas]{Current fluctuations in an interacting\\ active lattice gas} 

 \author{Stephy Jose}
\address{Tata Institute of Fundamental Research, Hyderabad, India, 500046}
\eads{\mailto{stephyjose@tifrh.res.in}}
\author{Rahul Dandekar}
\address{Institut de Physique Theorique, CEA, CNRS,\\ Universite Paris–Saclay, F–91191 Gif-sur-Yvette Cedex, France}
\eads{\mailto{rahul.dandekar@ipht.fr}}
\author{Kabir Ramola}
\address{Tata Institute of Fundamental Research, Hyderabad, India, 500046}
\eads{\mailto{kramola@tifrh.res.in}}
\date{\today}
\begin{abstract}
We study the fluctuations of the integrated density current across the origin up to time $T$ in a lattice model of active particles with hard-core interactions. This model is amenable to an exact description within a fluctuating hydrodynamics framework. We focus on quenched initial conditions for both the density and the magnetization fields and derive expressions for the cumulants of the density current, which can be matched with direct numerical simulations of the microscopic lattice model. For the case of uniform initial profiles, we show that the variance of the integrated current displays three regimes: an initial $\sqrt{T}$ rise with a coefficient given by the symmetric simple exclusion process, a cross-over regime where the effects of activity increase the fluctuations, and a large time $\sqrt{T}$ behavior with a prefactor which depends on the initial conditions, the Péclet number and the mean density of particles. Additionally, we study the limit of zero diffusion where the fluctuations intriguingly exhibit a $T^2$ behavior at short times. However, at large times, the fluctuations still grow as $\sqrt{T}$, with a coefficient that can be calculated explicitly. For low densities, we show that this coefficient can be expressed in terms of the effective diffusion constant $D_{\text{eff}}$ for non-interacting active particles.

\end{abstract}


\newpage
{\pagestyle{plain}
 \tableofcontents
\cleardoublepage}
\section{Introduction}

Active systems comprising particles that can self-propel and perform directed motion for intervals of time constitute a major class of non-equilibrium systems~\cite{vicsek1995novel,czirok1999collective,tailleur2008statistical,cavagna2010scale,cates2012diffusive,ramaswamy2010mechanics}. The steady states of active particle systems do not satisfy the principle of detailed balance because energy is dissipated at the microscopic scale in the bulk. Active systems have long attracted a lot of attention due to their biological relevance and applications in synthetic materials and soft matter industries. There have been numerous studies on the dynamics of active particles in different spatial dimensions using different microscopic models such as the run and tumble (RTP) model~\cite{evans2018run,malakar2018steady,mori2020universal,mori2020universalp,angelani2014first,martens2012probability} and the active Brownian motion (ABM) model~\cite{lindner2008diffusion,basu2018active,kumar2020active,romanczuk2010collective,romanczuk2012active}. Even though the literature of analytic results on active particle systems is vast, most of these studies focus on single particle models. 
There have been numerous studies on computing various quantities related to a single active particle such as the position distributions, first passage times, and large deviation functions. Given the rich behavior of active particles, there have also been numerous attempts to study active systems in different geometries, confining potentials, and with space-dependent activity~\cite{das2018confined,caprini2019active,sevilla2019stationary}. Even at the single particle level, active systems exhibit intriguing features such as non-Boltzmannian steady-state distribution, unusual first passage properties, and large deviation functions with different cross-over regimes~\cite{jose2022active,jose2022first,le2019noncrossing}.

Most analytical studies on active matter at the multi-particle level are based on fluid dynamic approaches, mean field theories, and gradient expansions~\cite{peshkov2012nonlinear,wittkowski2014scalar,nardini2017entropy,solon2018generalized,dandekar2020hard}. These models have been very successful in predicting many collective properties exhibited by active matter such as clustering, synchronous dynamics, motility induced phase separation (MIPS), etc. However, the lack of microscopic models that are amenable to an exact analysis has been an open problem within this field. Recently, an active lattice gas model with interactions has been introduced by Kourbane-Houssene et al~\cite{kourbane2018exact}. This model consists of hard core active particles on a lattice, and can be described by exact hydrodynamic equations that predict emergent behavior akin to MIPS. This framework can therefore be used to study the dynamics of interacting active particles, and corroborate the predictions of the hydrodynamic theory with microscopic simulations. Recent studies have extended this framework to a fluctuating hydrodynamics description~\cite{agranov2021exact}, as well as a macroscopic fluctuation theory (MFT)~\cite{agranov2023macroscopic,agranov2022entropy} which accounts for the Poissonian noise arising due to the flipping of the velocities of the particles. This allows for an investigation of several interesting aspects such as diverging correlation lengths, dynamical correlation functions, current fluctuations, large deviation functions, as well as entropy production~\cite{agranov2023macroscopic,agranov2022entropy}.

Macroscopic fluctuation theory~\cite{bertini2005current,bertini2006non,bertini2007stochastic,bertini2009towards} is a framework for investigating the fluctuating hydrodynamics of many-particle systems in the limit where the noise is small. In this limit, the evolution equations for stochastic systems take the form of a classical Hamiltonian field theory which can be used to compute the fluctuations of macroscopic variables and currents. MFT involves coupled partial differential equations for macroscopic observables and the associated conjugate fields and has been used to compute the large deviation functions associated with many quantities such as the densities of mass, charge, and energy, as well as their associated currents. 

A quantity of central interest that can be studied within the MFT is the integrated density current $Q_{\rho}(T)$ across the origin up to time $T$ for different diffusive processes. The time-dependence of the mean current, as well as its higher cumulants provides sensitive information regarding the large scale behavior of diffusive systems. Some examples of these processes that have been extensively studied include a collection of independent random walkers, the symmetric simple exclusion process (SSEP), and the KMP model in one dimension. SSEP is a model of hard core particles on a lattice, where the particles can hop symmetrically to any neighboring site if it is empty~\cite{levitt1973dynamics,arratia1983motion}. We note that the active lattice gas model studied in this paper reduces to the SSEP when the self-propulsion (activity) is set to zero. In the KMP model, a chain of mechanically uncoupled harmonic oscillators in one dimension is considered, where energy is redistributed between neighboring oscillators stochastically, while the total energy remains constant~\cite{kipnis1982heat}. Using MFT, it was shown that the cumulants of the integrated current for the above processes in one dimension scale as $\sqrt{T}$ for all macroscopic times $T$~\cite{krapivsky2012fluctuations}. Additionally, there have also been detailed studies on the differences between annealed versus quenched averaging for these models. In the annealed case, the initial positions of the particles are allowed to fluctuate, while in the quenched average, the particle positions are initially fixed. Interestingly, it was demonstrated that the prefactor of the current fluctuations for the annealed and quenched averages differs even as $T \rightarrow \infty$, suggesting the existence of long-term memory of initial conditions~\cite{derrida2009current2,derrida2009current,krapivsky2012fluctuations,banerjee2022role}. For these models with one conserved field, a perturbative approach to solve the MFT equations about the noiseless solution was introduced in \cite{krapivsky2012fluctuations}. This approach yields successive cumulants of the quantity being studied, in this case, the current across the origin at each order. Subsequently, this was used to obtain a general expression for the variance of the current, starting from a flat initial condition \cite{krapivsky2015tagged}, and to derive expressions for the higher-order cumulants \cite{krapivsky2014large}. Recently, the MFT equations for the SSEP have been exactly solved for annealed initial conditions \cite{mallick2022exact}, and the perturbative MFT approach has also been extended to long-ranged interactions \cite{dandekar2022dynamical}. 


Although many predictions of fluctuating hydrodynamics for this interacting active particle model have been tested against microscopic simulations, including dynamical correlation functions \cite{agranov2021exact}, other quantities remain inaccessible within this framework. A prime example is the fluctuations of the time-integrated current, which can be extracted in a systematic manner, using the macroscopic fluctuation theory. For annealed initial conditions, it has been shown previously~\cite{banerjee2020current} using exact techniques involving single-particle Green's functions that the variance of the integrated density current across the origin for non-interacting RTPs in one dimension exhibits different scaling behaviors in time. This is unlike the SSEP case where the variance of the integrated current grows as $\sqrt{T}$ at all times. In this work, we extend the recently developed MFT framework to study the integrated density current fluctuations for the interacting active particle model studied in~\cite{agranov2023macroscopic,agranov2022entropy}, but for quenched initial conditions. Interestingly, we find that the variance of the integrated current of the interacting active lattice gas with diffusion exhibits three regimes; a short-time regime where the variance grows as $\sqrt{T}$ which can be described by the SSEP, a cross-over regime, and a large time regime where the variance grows again as $\sqrt{T}$. We also study the case of zero diffusion where the system exhibits a new regime of behavior at short times, where the variance increases as $T^2$. We also demonstrate that the large time behavior of the current fluctuations of the interacting active gas is $\sqrt{T}$ with a prefactor that, for low densities, coincides with the effective diffusion constant $D_{\text{eff}}$ for noninteracting active walkers. 
We explicitly calculate the coefficient of the variance for finite densities as well, when interactions are important.

This paper is organized as follows. In section~\ref{sec:model}, we introduce the microscopic model used in the study. In sections~\ref{sec:hydrodynamics} and~\ref{sec:fluctuating hydrodynamics}, we explain the hydrodynamics and fluctuating hydrodynamics framework for this model developed in previous studies for completeness. We extend the MFT framework introduced in~\cite{agranov2023macroscopic} to compute the integrated current fluctuations in sections~\ref{sec:MFT} and~\ref{sec:int current}. We provide the details of the perturbative techniques used in the study and the expressions for the cumulants of the integrated current for general initial conditions in section~\ref{sec_perturbative}. In section~\ref{sec:interacting_gas}, we discuss our main results on the fluctuations of the integrated density current of the active lattice gas model for flat initial conditions. We also analyze the active gas model with zero diffusion in a separate subsection. We present the conclusions from the study in section~\ref{sec:conclusion}. Finally, we provide the details pertaining to some of the calculations and the microscopic as well as macroscopic simulations in appendices~\ref{appendix_alternate_exp}-\ref{appendix_simulations}.

\section{Microscopic model}
\label{sec:model}
\begin{figure} [!t]
\begin{center}
 \includegraphics[width=0.9\linewidth]{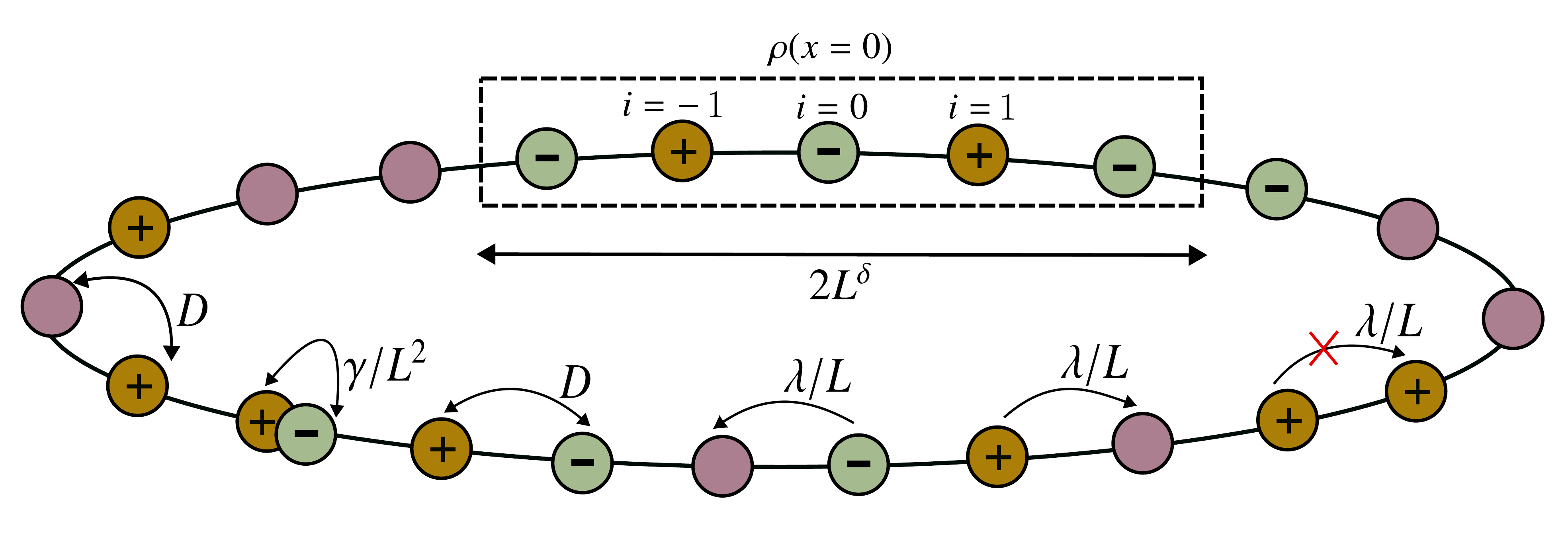}
\caption{ Lattice model of interacting active particles with different probability rates. (i) Neighboring sites exchange their occupancies at a diffusive rate $D$. A diffusive exchange can occur even if both the neighboring sites are occupied and is independent of the bias states of the particles. (ii) A $+$ particle can jump to the right neighboring site if that site is empty with a bias rate $\lambda/L$. Similarly, a $-$ particle can jump to the left neighboring site if that site is empty with a bias rate $\lambda/L$. Bias moves require one of the neighboring sites to be empty. (iii) Also, particles switch their states with a flipping rate $\gamma/L^2$. Disallowed transitions are represented by a red cross. The coarse-graining box is of size $2 L^\delta$ where $0<\delta<1$.}
\label{lattice_model}
\end{center}
\end{figure}
We consider a one-dimensional lattice bounded between $\left( -L/2, L/2\right]$ with interacting active particles where each site $i$ can be occupied by at most one particle~\cite{kourbane2018exact}. The lattice is periodic and is of size $L$. Each particle can be associated with an internal state $+$ or $-$ depending on the bias direction. The occupancies of site $i$ are denoted by the indicator variables $\mu_i^+$ and $\mu_i^-$. If the site $i$ is occupied by a $+$ particle, $\mu_i^+=1$ and $\mu_i^-=0$. Similarly, if the site $i$ is occupied by a $-$ particle, $\mu_i^-=1$ and $\mu_i^+=0$. Both $\mu_i^-=\mu_i^+=0$ if the site $i$ is empty. The dynamics take place according to the following microscopic rules

 (1) Neighboring sites exchange their occupancies at a diffusive rate $D$.
 
 (2) A $+$ particle can jump to the right neighboring site if that site is empty with a bias rate $\lambda/L$. Similarly, a $-$ particle can jump to the left neighboring site if that site is empty with a bias rate $\lambda/L$.
 
 (3) Particles switch their states with a flipping rate $\gamma/L^2$.
 
 The scalings of the bias and flipping rates with the system size $L$ ensure that all three processes contribute equally in the coarse-grained hydrodynamic regime~\cite{kourbane2018exact,agranov2021exact}. Figure~\ref{lattice_model} provides the schematic representation of the lattice sites with a few particles and the associated probability rates.

In this work we are concerned with the current fluctuations through the origin, starting from a quenched uniform initial condition where both the positions and states ($+$ and $-$) of the particles are fixed initially. This situation can be created in our microscopic simulations, with a fixed assignment of $+$ and $-$ particles. In particular, we study the initial condition with zero magnetization and uniform particle density where the hydrodynamic equations can be solved exactly.

\section{Hydrodynamic equations}
\label{sec:hydrodynamics}
Using the diffusive rescaling of space and time $x \rightarrow i/L$ and $t \rightarrow t/L^2$, one can define the coarse-grained plus and minus density fields $\rho^+(x,t)$ and $\rho^-(x,t)$ as
\begin{eqnarray}
\rho^+(x, t) &=&\frac{1}{2 L^{\delta}} \sum_{|i-L x|<L^{\delta}}\mu_{i}^{+},\nonumber\\
\rho^-(x, t) &=&\frac{1}{2 L^{\delta}} \sum_{|i-L x|<L^{\delta}}\mu_{i}^{-},
\end{eqnarray}
where the coarse graining parameter $\delta \in (0,1)$.
The hydrodynamic equations obeyed by the system were shown in \cite{kourbane2018exact} to be 
\begin{eqnarray}
	\partial_t\rho^+&=&D\partial^2_x\rho^+-\lambda\partial_x\left[\rho^+(1-\rho)\right]+\gamma(\rho^--\rho^+),\nonumber\\
	\partial_t\rho^-&=&D\partial^2_x\rho^-+\lambda\partial_x\left[\rho^-(1-\rho)\right]+\gamma(\rho^+-\rho^-).\label{eq:plus_minus}
\end{eqnarray}
In terms of the total density $\rho=\rho^++\rho^-$ and magnetization $m=\rho^+-\rho^-$ fields, the hydrodynamic equations can be rewritten as
\begin{eqnarray}
\partial_t \rho&=& D \partial^2_{x}\rho-\lambda\partial_x[m(1-\rho)],\nonumber\\
\partial_t m&=& D \partial^2_{x}m-\lambda\partial_x[\rho(1-\rho)]-2\gamma m. \label{rho_m_eq}
\end{eqnarray}

\section{Fluctuating hydrodynamics framework} 
\label{sec:fluctuating hydrodynamics}
The fluctuating hydrodynamic equations~\cite{agranov2021exact} obeyed by $\rho^+(x,t)$ and $\rho^-(x,t)$ can be derived as
\begin{eqnarray}
	\partial_t\rho^+&=&D\partial^2_x\rho^+-\lambda\partial_x\left[\rho^+(1-\rho)\right]+\gamma(\rho^--\rho^+)+\frac{ \sqrt{D}}{\sqrt{L}}\,\partial_x\eta^++\frac{\sqrt{\gamma}}{\sqrt{L}} \,\eta_K,\nonumber\\
	\partial_t\rho^-&=&D\partial^2_x\rho^-+\lambda\partial_x\left[\rho^-(1-\rho)\right]+\gamma(\rho^+-\rho^-)+\frac{\sqrt{D}}{\sqrt{L}} \,\partial_x\eta^--\frac{\sqrt{\gamma}}{\sqrt{L}} \,\eta_K.\label{eq:plus_minus_hdro}
\end{eqnarray}
The noises $\eta^+$ and $\eta^-$ are Gaussian noises with mean zero and delta correlations. 
These fluctuations are conjectured through an exact mapping with the ABC model~\cite{clincy2003phase,bodineau2011phase}.
However, the noise $\eta_K$ coming from tumbling events follows Poissonian statistics~\cite{agranov2023macroscopic}. They
come from slow local tumbling events of the $+$ and the $-$ particles. In the  ABC model, each site is occupied by a particle of type A, B, or C. The mapping follows by identifying $+$ particles,~$-$ particles, and holes in the active gas model with A, B, and C particles respectively in the ABC model~\cite{agranov2021exact}. This allows us to derive the correlations for the noise terms associated with the conservative fluxes in the hydrodynamic equations for the density fields of $+$ and $-$ particles as
\begin{eqnarray}
\langle\eta^{\pm} (x,t)\eta^{\pm} (x',t' )\rangle=2\rho^{\pm} (1-\rho^{\pm})\,\delta (x-x')\delta (t-t'),\nonumber\\ \langle\eta^+ (x,t)\eta^- (x',t' )\rangle=\langle\eta^- (x,t)\eta^+ (x',t' )\rangle=-2\rho^+\rho^-\delta (x-x')\delta (t-t').\label{ABC_correlations}
\end{eqnarray}

In terms of the total density and magnetization fields $\rho(x,t)$ and $m(x,t)$, the fluctuating hydrodynamic equations provided in equation~\eqref{eq:plus_minus_hdro} can be written as
\begin{eqnarray}
\partial_t \rho&=& D \partial^2_{x}\rho-\lambda\partial_x[m(1-\rho)]+\frac{\sqrt{D}}{\sqrt{L}} \,\partial_x\eta_\rho,\nonumber\\
\partial_t m&=& D \partial^2_{x}m-\lambda\partial_x[\rho(1-\rho)]-2\gamma m+\frac{1}{\sqrt{L}}\Big( \sqrt{D} \,\partial_x\eta_m+2\sqrt{\gamma}\,\eta_K \Big) . \label{eq:fluctuating_hydro}
\end{eqnarray}
The noise terms $\eta_{\rho}$ and $\eta_m$ are simply given as $\eta^{+}+\eta^{-}$ and $\eta^{+}-\eta^{-}$ respectively with mean zero and following correlations,
\begin{eqnarray}
\langle\eta_{\rho} (x,t)\eta_{\rho} (x',t')\rangle&=&\sigma_{\rho}\,\delta (x-x')\delta (t-t'),\nonumber\\
\langle\eta_{m} (x,t)\eta_{m} (x',t')\rangle&=&\sigma_{m}\delta (x-x')\delta (t-t'),\nonumber \\
\langle\eta_{\rho} (x,t)\eta_{m} (x',t')\rangle&=&\langle\eta_{m} (x,t)\eta_{\rho} (x',t')\rangle=\sigma_{\rho,m}\,\delta (x-x')\delta (t-t'),
\label{correlations}
\end{eqnarray}
and the noise amplitudes are given as
\begin{equation}
\sigma_{\rho}=2 \rho(1-\rho) ,\quad \sigma_{m}=2\left(\rho-m^{2}\right), \quad \sigma_{\rho, m}=2 m(1-\rho) .
\label{sigma}
\end{equation}

Using a second rescaling $t\rightarrow t \gamma$ and $x\rightarrow x \ell_s$ where $\ell_s=\sqrt{\gamma/D}$, the fluctuating hydrodynamic equations provided in equation~\eqref{eq:fluctuating_hydro} can be converted to the dimensionless form 
\begin{eqnarray}
\partial_{t} \rho &=-\partial_{x} J_{\rho}=-\partial_{x} \bar{J}_{\rho} +\sqrt{\frac{\ell_s}{L}}\partial_{x} \eta_{\rho}, \nonumber\\
\partial_{t} m &=-\partial_{x} J_{m}-2 K=-\partial_{x} \bar{J}_{m}-2 \bar{K} +\sqrt{\frac{\ell_s}{L}}(\partial_{x} \eta_{m}+2 \eta_K),
\label{fluctuating_hydrodynamics}
\end{eqnarray}
where the deterministic hydrodynamic components are given as
\begin{equation}
\bar{J}_{\rho}=-\partial_{x} \rho+\operatorname{Pe} m(1-\rho), \quad \bar{J}_{m}=-\partial_{x} m+\operatorname{Pe} \rho(1-\rho), \quad \bar{K}=m,
\label{currents_J}
\end{equation}
and the noise correlations are given by equation~\eqref{correlations}. Here, the  Péclet number $\mathrm{Pe}=\lambda / \sqrt{\gamma D}$ gives the ratio of the distance traveled by the particle due to pure bias to that by pure diffusion between two consecutive tumbles  and the field $K$ measures the local difference between the flips of $+$ particles to $-$ particles and $-$ particles to $+$ particles. The fluctuations of $K$ around its mean value $\bar{K}$ follow a Poisson distribution. The noise caused by these stochastic events can be straightforwardly accounted for in MFT, as demonstrated in the next section. In deriving equation~\eqref{fluctuating_hydrodynamics}, we have performed two rescalings of space and time coordinates. The first rescaling is the diffusive rescaling ($x\rightarrow i/L$, $t\rightarrow t/L^2$) where we rescale the space and time coordinates by the actual system size $L$. This results in $1/\sqrt{L}$ scaling of the noise terms. The second rescaling $t\rightarrow t \gamma$ and $x\rightarrow x \ell_s$ introduces an additional $\sqrt{\ell_s}$ scaling of the noise terms. We emphasize that the diffusive scaling has been performed in all calculations and plots presented in the subsequent sections, whereas in some subsequent calculations, we find it convenient to perform the second rescaling. In this case, we explicitly refer to these as ``rescaled'' equations.

The noiseless hydrodynamic equations in the rescaled coordinates can be obtained by setting the noise terms to zero in equation~\eqref{fluctuating_hydrodynamics}. These are simply given as
\begin{eqnarray}
\partial_{t} \rho &=-\partial_{x} \bar{J}_{\rho}, \nonumber\\
\partial_{t} m &=-\partial_{x} \bar{J}_{m}-2 \bar{K} ,
\label{hydrodynamics}
\end{eqnarray}
with the expressions for the deterministic currents provided in equation~\eqref{currents_J}.
We focus on the $\ell_s =\sqrt{\gamma/D}\rightarrow \infty$ limit in our analytical calculations presented in section~\ref{sec:interacting_gas}. This is the same limit studied in~\cite{kourbane2018exact,agranov2021exact} to derive the phase diagram associated with the interacting active gas model. The homogeneous steady state solutions to the hydrodynamic equations provided in equation~\eqref{hydrodynamics} are given as $\rho(x,t)=\bar{\rho},~m(x,t)=0$, where $\bar{\rho}$ is the mean density. It can be shown that the constant density solution $\rho (x,t)=\bar{\rho}$ and the zero magnetization solution $m (x,t)=0$ are linearly unstable when $\text{Pe}^2 (1-\bar{\rho})(2 \bar{\rho}-1)>2$. This defines the spinodal region of the system. In this region, a coexistence of dilute and dense phases was observed in~\cite{kourbane2018exact} and this is a characteristic feature of motility-induced phase separation (MIPS). We limit our study to the linearly stable region outside the spinodal curve in the phase diagram. This is given as $\text{Pe}^2 (1-\bar{\rho})(2 \bar{\rho}-1)<2$. In this parameter regime, the homogeneous solutions to the hydrodynamic equations are linearly stable.

\section{Macroscopic fluctuation theory framework}
\label{sec:MFT}

The fluctuating hydrodynamic equations provided in equation~\eqref{fluctuating_hydrodynamics} can be interpreted within the framework of macroscopic fluctuation theory, which allows for the direct computation of various quantities including the cumulants of the integrated current. The probability $P$ of observing a history of fields $\rho(x, t),~m(x, t),~J_{\rho}(x, t),~J_{m}(x, t),~K(x, t)$ in the rescaled space interval $-{\ell_s}/{2}<x \le{\ell_s}/{2}$ and rescaled time interval $0<t<T$ can be written as
\begin{equation}
-\ln P\left[\rho, m, J_{ \rho}, J_{ m}, K\right] = L \ell_{s}^{-1} {\mathcal{S}}\left[\rho, m, J_{ \rho}, J_{ m}, K\right].
\end{equation}
The prefactor ${L\ell_{s}}^{-1}$ appearing in the rhs of the above equation results from the rescalings of the space and time coordinates. 

The action ${\mathcal{S}}$ in the rescaled coordinates is given as
\begin{equation}
{\mathcal{S}}\left[\rho, m, J_{ \rho}, J_{ m}, K\right] =\int_{0}^{T} d t \int_{-\frac{\ell_s}{2}}^{\frac{\ell_s}{2}} d x\left(\mathcal{L}_{J}+\mathcal{L}_{K}\right) .
\label{S_func}
\end{equation}
where $\mathcal{L}_{J}$ accounts for the Gaussian current fluctuations due to the hops of the particles and $\mathcal{L}_{K}$ accounts for the Poisson tumble statistics. These Lagrangian densities have the explicit forms
\begin{equation}
 \mathcal{L}_{J}=\frac{1}{2}\left[\begin{array}{c}
J_{\rho}-\bar{J}_{\rho} \\
J_{m}-\bar{J}_{m}
\end{array}\right]^{\mathrm{T}} \mathbb{C}^{-1}\left[\begin{array}{c}
J_{\rho}-\bar{J}_{\rho} \\
J_{m}-\bar{J}_{m}
\end{array}\right],   
\end{equation}
where the correlation matrix $\mathbb{C}$ is given as
\begin{equation}
\mathbb{C}=\left[\begin{array}{cc}
\sigma_{\rho} & \sigma_{\rho, m} \\
\sigma_{\rho, m} & \sigma_{m}
\end{array}\right],  
\label{correlation_matrix}
\end{equation}
and
\begin{equation}
 \mathcal{L}_{K}=\rho-\sqrt{K^{2}+\left(\rho^{2}-m^{2}\right)}+K \ln \left[\frac{\sqrt{K^{2}+\left(\rho^{2}-m^{2}\right)}+K}{(\rho+m)}\right] .   
\end{equation}
This exact expression for $ \mathcal{L}_{K}$ has been derived in~\cite{agranov2023macroscopic} by considering the underlying Poisson process for the tumble events. 
For typical small fluctuations, the noise arising due to the flipping of states can be approximated as a Gaussian noise, and the above expression for the Lagrangian density reduces to $\mathcal{L}_{K}={{(K-m)}^2}/{\left(2\sigma_{K}\right)}$, where $\sigma_K=\rho$. It is sufficient to consider the action with Gaussian noises to compute the cumulants up to the second order. However, in this paper, we have considered the full large deviation form for the generality of the calculations.

\section{Time integrated current}
\label{sec:int current}

The macroscopic fluctuation theory of the active lattice gas has been used in~\cite{agranov2023macroscopic} to investigate the large deviation function of the density current averaged over the whole system. They showed that this large deviation function displays a dynamical phase transition between a stationary profile and traveling waves. While the average current is a global quantity, in this paper we investigate a local quantity, which is the total number of particles transferred through a single bond in the system. The fluctuations of this quantity have also been studied in other stochastic particle systems~\cite{derrida2009current2,dandekar2022macroscopic,banerjee2020current}, as they are related to the tagged particle fluctuations, which are also of enormous interest. We study the change in the mass in the right half of the system, 
\begin{equation}
\begin{aligned}
Q_\rho (T)&=&  \frac{1}{2}\int_{0}^{\frac{\ell_s}{2}} dx  \  [\rho(x,T) - \rho(x,0) ] \,.
\label{Qrt}
\end{aligned}
\end{equation}
We have,
\begin{equation}
\int_{0}^{\frac{\ell_s}{2}} dx  \  [\rho(x,T) - \rho(x,0) ] =
\int_0^T dt \int_{0}^{\frac{\ell_s}{2}} dx  \  [-\partial_x J_{\rho} ]=\int_0^T dt~  [ J_\rho(0,t)-J_\rho({\ell_s}/{2},t) ].
\end{equation}
This is the difference between the integrated current at
two opposite edges along the ring. 
The distribution of the
integrated current is symmetric and has a zero mean. Thus at short times, the quantity $Q_{\rho}(T)$ measures the integrated current across the
origin in an infinite system. On a lattice of size $L$, finite size effects set in at microscopic times of order $\sim L^2$, and the fluctuations of this quantity saturate after such a time.

We next compute the moment generating function for the integrated current via the MFT formulation,
\begin{equation}
  \left\langle e^{L\ell_s^{-1}\Lambda Q_{\rho}(T)} \right\rangle= \int {\cal D} \rho{\cal D} m {\cal D} J_\rho {\cal D} J_m {\cal D} K \, e^{ L\ell_s^{-1} \left(\Lambda Q_{\rho}(t)-{{\cal S}}\right)}
    \, \prod_{x,t} \delta (\dot{\rho} + \partial_x J_\rho) 
    \,\delta ( \dot{m} + \partial_x J_m + 2K) \,,
    \label{path_integral}
\end{equation}
where the action $\cal S$ is defined in equation~\eqref{S_func}.
The moment-generating function encodes the full statistics of the integrated current $Q_{\rho}(T)$. The logarithm of the moment generating function yields the cumulant generating function from which the cumulants can be extracted by collecting terms that appear at the same powers of $\Lambda$.
The Dirac delta functions appearing in the above equations ensure that the continuity equations in equation~\eqref{fluctuating_hydrodynamics} for the density and magnetization fields are satisfied at each point of space and time ($x,t$).
We can use the integral representation of the delta functions, where we introduce the auxiliary fields, ${p}_\rho(x,t)$ and ${p}_m(x,t)$ which are also periodic. Equation~\eqref{path_integral} thus translates to
\begin{eqnarray}
  \left\langle e^{L\ell_s^{-1}\Lambda Q_{\rho}(T)} \right\rangle&=& \int {\cal D} \rho{\cal D} m {\cal D} J_\rho {\cal D} J_m {\cal D} K {\cal D} p_\rho {\cal D} p_m \, \exp \Bigg[ L\ell_s^{-1} \Big(\Lambda Q_{\rho}(t)-{{\cal S}}\nonumber\\&
   +&\iint dx dt\left( p_\rho \left( \dot{\rho} + \partial_x J_\rho \right)+p_m 
    \left( \dot{m} + \partial_x J_m + 2K \right) \right)\Big)\Bigg].   \label{path_integral2}
\end{eqnarray}
In the limit $L\ell_s^{-1} \xrightarrow[]{} \infty$, we arrive at a large deviation form for the cumulant generating function with 
\begin{equation}
 \frac{1}{L\ell_s^{-1}} \ln\left\langle e^{L\ell_s^{-1}\Lambda Q_{\rho}(T)} \right\rangle  = \psi_{\rho}(\Lambda,T),
 \label{large_dev_form}
\end{equation}
and $\psi_{\rho}(\Lambda,T)$ is the large deviation free energy function or the scaled cumulant generating function. In this limit, $\psi_{\rho}(\Lambda,T)$ can be obtained by a saddle point evaluation of the integral in equation~\eqref{path_integral2}. The MFT equations presented in this paper are generic and hold for any value of $\ell_s$ subject to the condition that $L\ell_s^{-1} \rightarrow \infty$. However, in the calculations presented in section~\ref{sec:interacting_gas}, we choose the particular limit $L \xrightarrow{} \infty,~\ell_s \xrightarrow[]{} \infty$, with $L\ell_s^{-1} \rightarrow \infty$. This is because $\ell_s \xrightarrow[]{} \infty$ is an easy limit to consider in the analytical computations presented in section~\ref{sec:interacting_gas}. 

After integrating over $J$ and $K$ fields in equation~\eqref{path_integral2}, the large deviation function $\psi_{\rho}(\Lambda,T)$ can be computed by maximizing an action ${\cal S}_\mathrm{tot}$ given as~\cite{agranov2023macroscopic}
\begin{equation}
 \psi_{\rho}(\Lambda,T)=\max_{\{\rho,m,p_\rho,p_m \}}\left[\Lambda  Q_{\rho}(T)-  {\cal S}_\mathrm{tot}\right],
\label{large_deviation_free_energy_function}
\end{equation}
and ${\cal S}_\mathrm{tot}$ can be derived as
\begin{equation}
    {\cal S}_\mathrm{tot} = \int_{-\frac{\ell_s}{2}}^{\frac{\ell_s}{2}} dx \int_0^T dt \, 
     \Big\{
     \dot{\rho} {p}_\rho + \dot{m} p_m - {\cal H} [\rho, m, {p}_\rho, p_m]  \Big\}.    \label{minimized_action}
\end{equation}
The Hamiltonian density ${\cal H}$ appearing in the above equation can be computed as~\cite{agranov2023macroscopic}
\begin{equation}
    {\cal H} [\rho, m, {p}_\rho, p_m] = \frac{1}{2} \left[ \begin{array}{c}
	    \partial_x {p}_\rho\\
	    \partial_x p_m
	\end{array} \right]^\mathrm{T} \mathbb{C}
	\left[ \begin{array}{c}
	    \partial_x {p}_\rho\\
	    \partial_x p_m
	\end{array} \right] + \bar{J}_\rho \partial_x {p}_\rho + \bar{J}_m \partial_x p_m + 2 \rho \sinh^2 p_m - m \sinh 2 p_m~.
\end{equation}
The term $\Lambda Q_\rho (T )$ appearing in equation~\eqref{large_deviation_free_energy_function} determines the temporal boundary conditions for the conjugate fields.
One can now determine the equations obeyed by the optimal trajectories by considering small variations of the fields $\rho \rightarrow \rho+\delta \rho,~m \rightarrow m+\delta m,~p_\rho \rightarrow p_\rho+\delta p_\rho$ and $p_m \rightarrow p_m+\delta p_m$.
This yields four bulk Hamiltonian MFT equations for the fields $\rho,~m,~p_{\rho}$ and $p_m$ at the optimum.

\begin{eqnarray}
\hspace{-2.3 cm}
 \partial_t \rho = \frac{\delta {\cal H}}{\delta p_\rho} = - \partial_x \left[ \sigma_\rho \partial_x p_\rho + \sigma_{\rho, m} \partial_x p_m + \bar{J}_\rho \right],\nonumber\\
 \hspace{-2.3 cm}
 \partial_t m = \frac{\delta {\cal H}}{\delta p_m} = - \partial_x \left[ \bar{J}_m + \sigma_m \partial_x p_m + \sigma_{\rho, m} \partial_x p_\rho \right] + 2 (\rho \sinh 2 p_m - m \cosh 2p_m), \nonumber\\
 \hspace{-2.3 cm}
 \partial_t p_\rho = - \frac{\delta {\cal H}}{\delta \rho} = - \partial_x^2 p_\rho - (1 - 2 \rho) {(\partial_x p_\rho)}^2 + 2 m \partial_x p_\rho \partial_x p_m  - {(\partial_x p_m)}^2  \nonumber\\
 + \text{Pe} m \partial_x p_\rho - \text{Pe} (1 - 2 \rho) \partial_x p_m - 2 {\sinh}^2 p_m,\nonumber\\
 \hspace{-2.3 cm}
 \partial_t p_m = - \frac{\delta {\cal H}}{\delta m} = - \partial_x^2 p_m + 2 m (\partial_x p_m)^2 - 2 (1 - \rho) \partial_x p_\rho \partial_x p_m 
 - \text{Pe} (1 - \rho) \partial_x p_\rho + \sinh 2 p_m.\nonumber\\
 \label{mft_eqns}
\end{eqnarray}

The SSEP limit can be obtained by setting the Péclet number Pe$=0$ in the above equations. This corresponds to the case where the particles are non-motile. When Pe equals zero, the solution for $p_m$ is zero, causing the equations for $\rho$ and $p_\rho$ to become decoupled from the equation for $m$. These decoupled equations are identical to those of the SSEP. The MFT equations for the SSEP~\cite{derrida2009current2} are thus obtained as
\begin{eqnarray}
 \partial_t \rho &=&   -  \partial_x \left[\sigma_\rho \partial_x p_{\rho} +\bar{J}_{\rho}\right],\nonumber\\
 \partial_t p_{\rho} &=&  - \left[\partial_x^2 p_{\rho}+(1-2 \rho){(\partial_x p_{\rho})}^2 \right],~\text{for SSEP},
 \label{mft_eqns_SSEP}
\end{eqnarray}
with $\bar{J}_\rho=-\partial_x \rho$.

We note that although the expressions in equation~\eqref{mft_eqns} represent the optimal path of the density and magnetization fields, the boundary conditions are yet to be determined~\cite{derrida2009current2,krapivsky2012fluctuations}. These are set by the quantities being measured. In this study, we focus on the time-integrated current up to a time $T$, which sets an initial condition on the $\rho$ and $m$ fields and a final condition on $p_{\rho}$ and $p_{m}$. These represent sufficient conditions to uniquely specify the trajectory, and therefore one can obtain predictions for the integrated current which can then be matched with microscopic simulations. Since we  study the integrated density current up to time $T$, we obtain the final time boundary conditions on the auxiliary fields as
\begin{equation}
  \left\lbrace \begin{array}{l}
  p_{\rho}(x,T) = \Lambda \theta (x),\\
  p_m(x,T) = 0.
  \end{array}\right.
  \label{boundary_cond_p}
\end{equation}

\section{Perturbative framework}
\label{sec_perturbative}
\begin{figure} [!t]
 \includegraphics[width=1.0\linewidth]{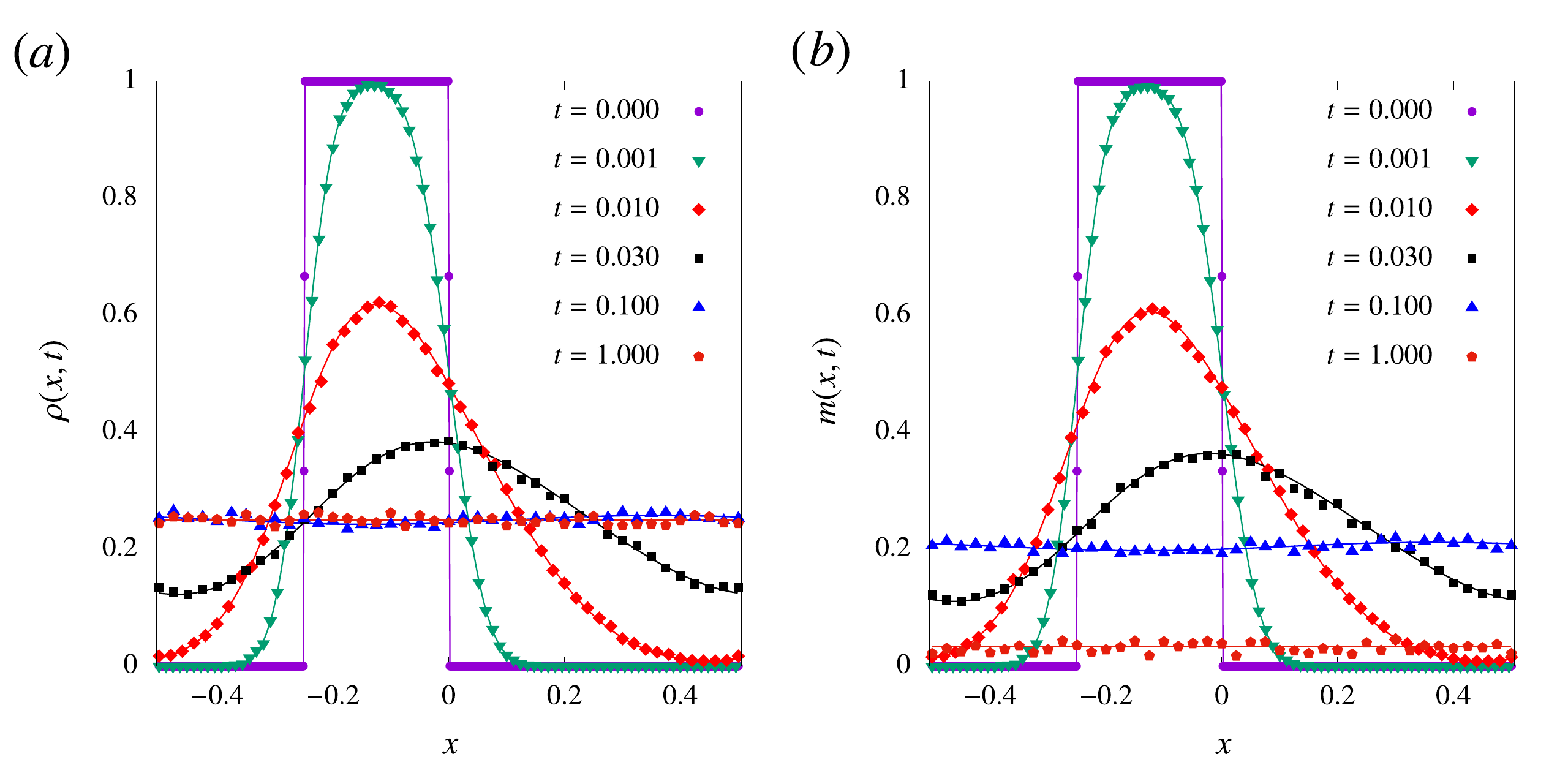}
\caption{Evolution of the (a) density $\rho(x,t)$ and (b) magnetization $m(x,t)$ fields starting from identical step initial conditions for fixed parameter values $D=1,~\lambda=10,~\gamma=1$. The points are obtained from Monte Carlo simulations and the solid curves are obtained from direct numerical integration of the zeroth order hydrodynamic equations provided in equation~(\ref{zeroth_order_eqns}) using finite difference methods. From the profiles, one can clearly observe the effects of advection along with the diffusive relaxation towards the uniform steady state.} In the microscopic simulations, we have used a lattice of size $L=1000$ with $250$ particles. The simulation data is averaged over $2000$ realizations. The rates as well as the spatial and temporal coordinates in the above plots have the diffusive scaling~($D \rightarrow D,~\lambda \rightarrow \lambda/L,~\gamma \rightarrow \gamma/L^2,~x \rightarrow i/L$,~ $t \rightarrow t/L^2$).
\label{fig:zeroth_order_fields}
\end{figure}

For small deviations from the average current, one can use the perturbative approach introduced in~\cite{krapivsky2012fluctuations} to compute the current fluctuations. We expand the fields $(\rho,~m,~p_{\rho},~p_m)$ about the solutions of the noiseless hydrodynamic equations $(\rho_0,~m_0,~0,~0)$ with $\Lambda$ as the perturbation parameter. Here, $\Lambda$ measures the noise strength and $\Lambda=0$ corresponds to the noiseless case. The fields are expanded as
\begin{eqnarray}
\rho &= \rho_0 + \Lambda \rho_1 + \Lambda^2 \rho_2+\ldots~,\nonumber\\
p_{\rho} &= \qquad \Lambda {p_{\rho}}_1 + \Lambda^2 {p_{\rho}}_2+\ldots~,\nonumber\\
m &= m_0 + \Lambda m_1 + \Lambda^2 m_2+\ldots~, \nonumber\\
p_{m} &= \qquad \Lambda {p_m}_1 + \Lambda^2 {p_m}_2+\ldots~.
\label{expansions}
\end{eqnarray}
Substituting these expressions into equation~(\ref{mft_eqns}) yields the zeroth order (in $\Lambda$) equations
\begin{eqnarray}
 \partial_t \rho_0 &=  - \partial_x   {\bar{J^0_\rho}} ,\nonumber\\
 \partial_t m_0 &=  - \partial_x {\bar{J^0_m}}  - 2m_0.
 \label{zeroth_order_eqns}
\end{eqnarray}
In the above equations, the superscript and subscript ``$0$" indicates that the corresponding fields are at zeroth order (noiseless). To zeroth order, we recover the noiseless hydrodynamic equations for the density and magnetization fields as in equation~\eqref{hydrodynamics}, with the $J$ fields defined in equation~\eqref{currents_J}. These zeroth-order equations have been shown to yield numerically exact results through a match with microscopic profiles and the numerical solutions of the coupled non-linear differential equations. Figure~{\ref{fig:zeroth_order_fields}} provides our results for the match between the numerical integration and the microscopic simulations. In this figure, we have used step initial profiles for both the density and magnetization fields. Substituting the expansions provided in equation~\eqref{expansions} into equation~(\ref{mft_eqns}) also yields the first-order (in $\Lambda$) equations
\begin{eqnarray}
\hspace{-1.75 cm}
 \partial_t {p_\rho}_1   = - \partial_x^2 {p_\rho}_1 + \text{Pe} m_0 \partial_x {p_\rho}_1 - \text{Pe} (1 - 2\rho_0) \partial_x {p_m}_1,\nonumber\\
 \hspace{-1.75 cm}
 \partial_t {p_m}_1 =  - \partial_x^2 {p_m}_1  - \text{Pe}  (1 - \rho_0) \partial_x {p_\rho}_1+2 {p_m}_1,\nonumber\\
 \hspace{-1.75 cm}
 \partial_t \rho_1 =  - \partial_x \big[-\partial_x \rho_1+   {\sigma^0_\rho}\partial_x {p_\rho}_1+{\sigma^0_{\rho,m}}\partial_x {p_m}_1  -\text{Pe}m_0\rho_1+\text{Pe}m_1(1 - \rho_0)\big],\nonumber\\
 \hspace{-1.75 cm}
 \partial_t m_1 =  - \partial_x \big[-\partial_x m_1+   \sigma^0_m\partial_x {p_m}_1+\sigma^0_{\rho,m}\partial_x {p_\rho}_1 +\text{Pe}\rho_1(1 - 2 \rho_0) \big]+4 \rho_0 {p_m}_1-2m_1.\nonumber\\
 \label{perturbative_eqns}
\end{eqnarray}

We next turn to the computation of the cumulants of the integrated current using the above perturbative equations. For the case of quenched average in density and magnetization fields, the initial conditions for $\rho$ and $m$ fields do not have any fluctuations. Thus we have
\begin{equation}
  \left\lbrace \begin{array}{l}
  {\rho}_0(x,0) = {\rho}(x,0),\\
  m_0(x,0) = m(x,0),
  \end{array}\right.\,
    \label{ic_zeroth}
\end{equation}
and
\begin{equation}
  \left\lbrace \begin{array}{l}
  {\rho}_1(x,0) = 0,\\
  m_1(x,0) = 0.
  \end{array}\right.\,
\end{equation}
Using equations~\eqref{expansions} and \eqref{boundary_cond_p}, the boundary conditions on the first-order conjugate fields translate to
\begin{equation}
  \left\lbrace \begin{array}{l}
  {p_\rho}_1(x,T) = \theta (x),\\
  {p_m}_1(x,T) = 0.
  \end{array}\right.\,
    \label{eq_current_ic1}
\end{equation}

We may also define an expansion of the integrated current using the expressions provided in equation~\eqref{expansions} as
\begin{eqnarray}
Q_{\rho}(T) =  Q_{\rho_0}(T) + \Lambda  Q_{\rho_1}(T) + ...
\end{eqnarray}
where the integrated currents up to the first order are given as
\begin{subequations}
\begin{align}
\label{eq:rho0}
 Q_{\rho_0}(T)  &= \int_{0}^{\frac{\ell_s}{2}} dx  \  [\rho_0(x,T) - \rho_0(x,0) ]~,\\
 \label{eq:rho1}
 Q_{\rho_1}(T)  &= \int_{0}^{\frac{\ell_s}{2}} dx  \  [\rho_1(x,T) ]~.
\end{align}
\label{currents}
\end{subequations}
We show in appendix~\ref{appendix_alternate_exp} that the above expressions directly reduce to the first and second cumulants of the integrated density current. Thus we obtain
\begin{equation}
  \langle Q_{\rho}(T) \rangle_c= Q_{\rho_0}(T),
\label{first_cumulant1}
\end{equation}
\begin{equation}
  \langle {Q_{\rho}(T)}^2 \rangle_c=   Q_{\rho_1}(T).
\label{second_cumulant1}
\end{equation}

We notice that the first cumulant depends only on the zeroth order field $\rho_0$ and the second cumulant depends only on the first order field $\rho_1$. Although the above expressions are the exact expressions for the first and second cumulants, the calculation of the second cumulant involves solving the $\rho_1$ field which in turn requires solving the six coupled equations provided in equations~\eqref{zeroth_order_eqns} and~\eqref{perturbative_eqns}.
We next derive alternate expressions for the cumulants from the perturbative expansions which require only solving the first two equations in~\eqref{perturbative_eqns} along with the hydrodynamic equations in~\eqref{zeroth_order_eqns}.
We substitute the expansions from equation~(\ref{expansions}) into equation~(\ref{large_deviation_free_energy_function}). To second order in $\Lambda$, we obtain
\begin{equation}
    \psi_{\rho}(\Lambda,T) = \Lambda  Q_{\rho_0}(T)+ \Lambda^2 Q_{\rho_1}(T)-  {\cal S}_\mathrm{tot}(\Lambda),
    \label{psi}
\end{equation}
where $ {\cal S}_\mathrm{tot}(\Lambda)$ is the expansion of the total action provided in equation~\eqref{minimized_action} up to second order in $\Lambda$ and the fields obey the MFT equations provided in equation~\eqref{mft_eqns}. Substituting the MFT equations provided in equation~\eqref{mft_eqns} into equation~\eqref{minimized_action} and integrating by parts, we obtain
\begin{eqnarray}
{\cal S}_\mathrm{tot} &=&  \int_{-\frac{\ell_s}{2}}^{\frac{\ell_s}{2}} dx \int_0^T dt~ \Bigg\{ (m+2 \rho p_m)\sinh 2p_m -m 2p_m\cosh 2 p_m\nonumber\\&&-2\rho \sinh^2p_m +\frac{1}{2} \left[ \begin{array}{c}
	    \partial_x {p_\rho}\\
	    \partial_x {p_m}
	\end{array} \right]^\mathrm{T} {\mathbb{C}}
	\left[ \begin{array}{c}
	    \partial_x {p_\rho}\\
	    \partial_x {p_m}
	\end{array} \right]\Bigg\}.
    \label{stot}
\end{eqnarray}
Expanding the above expression up to the second order in $\Lambda$, we obtain
\begin{eqnarray}
\hspace{-1 cm}
{\cal S}_\mathrm{tot}(\Lambda) =  \Lambda^2\int_{-\frac{\ell_s}{2}}^{\frac{\ell_s}{2}} dx \int_0^T dt~ \Bigg\{2\rho_0 {{p_m}_1}^2 +\frac{1}{2} \left[ \begin{array}{c}
	    \partial_x {p_\rho}_1\\
	    \partial_x {p_m}_1
	\end{array} \right]^\mathrm{T} {\mathbb{C}}_{\rho_0,m_0}
	\left[ \begin{array}{c}
	    \partial_x {p_\rho}_1\\
	    \partial_x {p_m}_1
	\end{array} \right]\Bigg\}.
    \label{stot_simplified}
\end{eqnarray}
Finally using equations~\eqref{psi} and~\eqref{stot_simplified}, we obtain the expression for the scaled cumulant generating function (up to second order in $\Lambda$) as
\begin{eqnarray}
    \psi_{\rho}(\Lambda,T) &=& \Lambda  Q_{\rho_0}(T)+ \Lambda^2 \Bigg\{ Q_{\rho_1}(T)-\int_{-\frac{\ell_s}{2}}^{\frac{\ell_s}{2}} dx \int_0^T dt \, \Bigg[2 \rho_0 {{p_m}_1}^2\nonumber\\&&+
     \frac{1}{2} \left[ \begin{array}{c}
	    \partial_x {p_\rho}_1\\
	    \partial_x {p_m}_1
	\end{array} \right]^\mathrm{T} {\mathbb{C}}_{\rho_0,m_0}
	\left[ \begin{array}{c}
	    \partial_x {p_\rho}_1\\
	    \partial_x {p_m}_1
	\end{array} \right]\Bigg]\Bigg\}.
\label{psi_perturbation}
\end{eqnarray}
The above expression is different from the symmetric simple exclusion process (SSEP) because of the additional $2 \rho_0 p_{m_1}^2$ term. By setting Pe$ =0$, which sets $p_m=0$, we can obtain the SSEP limit of the above expression. This yields
\begin{eqnarray}
\hspace{-1cm}
    {\psi_{\rho}(\Lambda,T)}_{\text{SSEP}} &=& \Lambda  Q_{\rho_0}(T)+ \Lambda^2 \left\{ Q_{\rho_1}(T)-\int_{-\frac{\ell_s}{2}}^{\frac{\ell_s}{2}} dx \int_0^T dt \, \left[
     \frac{1}{2} \sigma_{\rho_0}	    {(\partial_x {p_\rho}_1)}^2\right]\right\}.
\label{psi_perturbation_SSEP}
\end{eqnarray}
This is exactly the same expression derived in~\cite{krapivsky2012fluctuations}.

By definition, the cumulant generating function can also be expanded as
\begin{equation}
    \psi_{\rho}(\Lambda,T) = \Lambda  \langle Q_{\rho}(T) \rangle_c+ \frac{\Lambda^2}{2} \langle {Q_{\rho}(T)}^2 \rangle_c +\ldots~.
\label{psi_definition}
\end{equation} 
Collecting terms that appear at the same order in $\Lambda$ in equations~(\ref{psi_perturbation}) and (\ref{psi_definition}), we obtain the expression for the first cumulant of the integrated current as in equation~\eqref{first_cumulant1}. The expression for the second cumulant is now obtained as
\begin{equation}
  \langle {Q_{\rho}(T)}^2 \rangle_c=  2 Q_{\rho_1}(T)-\int_{-\frac{\ell_s}{2}}^{\frac{\ell_s}{2}} dx \int_0^T dt \, \Bigg \{4 \rho_0 {{p_m}_1}^2+
   \left[ \begin{array}{c}
	    \partial_x {p_\rho}_1\\
	    \partial_x {p_m}_1
	\end{array} \right]^\mathrm{T} {\mathbb{C}}_{\rho_0,m_0}
	\left[ \begin{array}{c}
	    \partial_x {p_\rho}_1\\
	    \partial_x {p_m}_1
	\end{array} \right]\Bigg \},
\label{second_cumulant}
\end{equation}
where $Q_{\rho_0}(T)$ and $Q_{\rho_1}(T)$ are defined in equation~\eqref{currents}.
We notice that the expression for the first cumulant depends only on the zeroth-order fields and the expression for the second cumulant depends only on the zeroth and first-order fields. The specific structure of the action for the system allows for exact cancellations of the fields appearing at higher orders.

The expression for the first cumulant obtained from both methods is exactly the same. Comparing the expressions for the second cumulant obtained using both the methods in equations~(\ref{second_cumulant}) and~(\ref{second_cumulant1}), we obtain
\begin{equation}
  \langle {Q_{\rho}(T)}^2 \rangle_c=   \int_{-\frac{\ell_s}{2}}^{\frac{\ell_s}{2}} dx \int_0^T dt \, \Bigg \{4 \rho_0 {{p_m}_1}^2+
   \left[ \begin{array}{c}
	    \partial_x {p_\rho}_1\\
	    \partial_x {p_m}_1
	\end{array} \right]^{\mathrm{T}} {\mathbb{C}}_{\rho_0,m_0}
	\left[ \begin{array}{c}
	    \partial_x {p_\rho}_1\\
	    \partial_x {p_m}_1
	\end{array} \right]\Bigg \}.
\label{second_cumulant2}
\end{equation}
We note from the above equation that the variance of the density current explicitly involves only the zeroth order fields and the first order momentum fields. This is similar to previous results obtained for passive particles~\cite{krapivsky2012fluctuations} where the current fluctuations are given by the second term of the above equation but with just one field (density). 
The active case, in contrast to the passive case, has the extra term $4 \rho_0 {{p_m}_1}^2$ which explicitly involves the square of the conjugate magnetization field. 
In practice, it is easier to integrate the $p$ fields at first order as it involves only the zeroth order fields and the first order $p$ fields themselves (see equation~\eqref{perturbative_eqns}). The advantage of the perturbation expansion is that we are able to provide analytic expressions for the $p$ fields as we show in the latter sections.
Equation~\eqref{second_cumulant2} represents one of the main results of our study. Given the solutions of the MFT equations \eqref{perturbative_eqns} up to first order, the cumulants of the current can be derived using the above expressions. 

We note that these are the typical fluctuations as we expand the solutions about the deterministic hydrodynamic solutions. To obtain the expression for the variance, we expand the action, retaining terms up to second order in the parameter $\Lambda$. This expansion allows to approximate the large deviation function, or the cumulant generating function, up to second order in $\Lambda$. This approximation provides insights into the small fluctuations of the current from its mean value. However, to account for large fluctuations, which are manifested in the tail of the large deviation function, it is necessary to include higher-order terms in $\Lambda$ in the perturbation expansion. The perturbative framework discussed in this section holds for arbitrary initial conditions for $\rho$ and $m$ fields. However, the computation of the fields and subsequently the integral in equation~\eqref{second_cumulant2} is in general difficult. In the next section, we discuss the case of uniform initial conditions for the $\rho$ and $m$ fields, where it is possible to compute exact analytical expressions for the current fluctuations.

\section{Current fluctuations for uniform initial conditions}
\label{sec:interacting_gas}
In this section, we use the perturbative framework developed in the previous section to predict the current fluctuations in an active lattice gas. The perturbative framework allows for the computation of the second cumulant from the first-order solutions of the $p_\rho$ and $p_m$ fields. However, since the equations governing the fields at each order are non-linear, finding closed-form solutions even at the zeroth order is challenging. We, therefore, turn to cases where the zeroth order fields can be exactly determined and therefore be used to provide solutions at the first order.
The simplest case for the active lattice gas is when the density fields $\rho^+$ and $\rho^-$ for the two types of particles are exactly identical. This corresponds to the case with uniform density and zero magnetization throughout the lattice.
We consider the initial condition
\begin{equation}
    \rho(x,0)=\bar{\rho},~m(x,0)=0.
    \label{init_fields}
\end{equation}
Although the above form of the initial condition defined within a finite region $-{\ell_s}/{2}<x\le{\ell_s}/{2} $ which respects periodic boundary conditions is easy to realize in numerical simulations, for our analytical studies we focus on the case of $\ell_s \xrightarrow{} \infty$. This is for the simplicity of the calculations presented in this section. For this, we first take the limit  $L \xrightarrow{} \infty$, then take the limit $\ell_s \xrightarrow{} \infty$ with $L\ell_s^{-1} \rightarrow \infty$ as described in section~\ref{sec:int current}. Therefore, our microscopic simulations for finite lattice size deviate from the theory after a certain (large) time once the boundary effects become important. We do not probe these boundary effects in our present work.

\subsection{Interacting active lattice gas with non-zero diffusion}
In this section, we study the current fluctuations in an interacting active lattice gas with a finite non-zero value of the diffusion constant.
For the homogeneous initial conditions equation~\eqref{init_fields}, the zeroth order MFT equations presented in equation~(\ref{zeroth_order_eqns}) admit analytical solutions of the form
\begin{equation}
\rho_0(x,t)=\bar{\rho},~m_0(x,t)=0.
\label{soln:zeroth_order_zero_mag}
\end{equation}
Substituting these solutions into the first-order equations for the conjugate fields in equation~(\ref{perturbative_eqns}) yields
\begin{eqnarray}
  \partial_t {p_\rho}_1   &=& - \partial_x^2 {p_\rho}_1  - \text{Pe} (1 - 2\bar{\rho}) \partial_x {p_m}_1,\nonumber\\
 \partial_t {p_m}_1 &=&  - \partial_x^2 {p_m}_1  - \text{Pe}  (1 - \bar{\rho}) \partial_x {p_\rho}_1+2 {p_m}_1.
\label{eq:zero_mag}
\end{eqnarray}
The above equations are in the rescaled coordinates (that is, $x \xrightarrow{} x \ell_s$ and $t \xrightarrow{} t \gamma$). Being linear, these equations can be solved exactly. For general initial conditions, the first-order equations involve nonlinear terms and are hard to solve analytically.
The above equations are to be solved with the time boundary conditions $p_{\rho_1}(x,T)=\theta(x)$ and $p_{m_1}(x,T)=0$. Using the transformation $\tau \rightarrow T-t$, these equations can be solved as an initial condition problem in the Fourier space. We define the Fourier transform of the field $p$ as $\tilde p(k,\tau)=\int_{-\infty}^{\infty}dx e^{-i k x} p (x,\tau)$ and the inverse Fourier transform as $p(x,\tau)=\frac{1}{2\pi}\int_{-\infty}^{\infty}dk e^{i k x}\tilde p (k,\tau)$. If $\ell_s$ is finite, the Fourier transform is defined as a discrete summation over modes rather than a continuous transform.
Taking a Fourier transform of equation~(\ref{eq:zero_mag}) yields the matrix equation
\begin{equation}
\frac{\partial}{\partial \tau}
\ket{\tilde p (k,\tau)}
=\mathcal{M}(k)
\ket{\tilde p (k,\tau)},
\label{eq:matrix}
\end{equation}
where the column vector $\ket{\tilde p (k,\tau)}$ is given as
\begin{equation}
\ket{\tilde p (k,\tau)}=       {\begin{pmatrix}
\tilde p_{\rho_1}(x,\tau)\\
\tilde p_{m_1}(x,\tau)
\end{pmatrix}}, 
\label{column_vec}
\end{equation}
and the matrix $\mathcal{M}(k)$ is given as
\begin{equation}
    \mathcal{M}(k)=\begin{pmatrix}
 -k^2 & -i k\text{Pe}(2\bar{\rho}-1) \\
i k \text{Pe}(1-\bar{\rho}) & -(k^2+2)
 \end{pmatrix}.
\end{equation}
Equation~(\ref{eq:matrix}) can be solved by diagonalizing the matrix $\mathcal{M}(k)$. 
The eigenvalues $\epsilon_1(k)$, $\epsilon_2(k)$ and the eigenvectors $\ket{\psi_1(k)}$, $\ket{\psi_2(k)}$ of the matrix $\mathcal{M}(k)$ are given as
\begin{equation}
\epsilon_1(k)=-1-k^2-\sqrt{1+k^2 g},~ \epsilon_2(k)=-1-k^2+\sqrt{1+k^2 g}~,  
\label{eigen_val}
\end{equation}
and
\begin{equation}
\ket{\psi_1(k)}=    {\begin{pmatrix}
\frac{i \left(-1+\sqrt{1+k^2 g}\right)}{k \text{Pe} (1-\bar{\rho})}\\
 1
\end{pmatrix}},~\ket{\psi_2(k)}=    {\begin{pmatrix}
\frac{i \left(-1-\sqrt{1+k^2 g}\right)}{k \text{Pe} (1-\bar{\rho})}\\
 1
\end{pmatrix}}.
\label{eigen_vec}
\end{equation}
The constant $g$ appearing in the above expressions has the explicit form
\begin{equation}
    g=\text{Pe}^2 (1-\bar{\rho})(2 \bar{\rho}-1).
\label{g}
\end{equation}
We note that the above constant $g$ is the same factor that appears in the equation of the spinodal curve and the correlation length in~\cite{tailleur2008statistical,agranov2021exact}.
We have the initial condition for the conjugate fields as
\begin{equation}
    {\begin{pmatrix}
\tilde p_{\rho_1}(x,\tau=0)\\
\tilde p_{m_1}(x,\tau=0)
\end{pmatrix}}=
    {\begin{pmatrix}
\theta(x)\\
 0
\end{pmatrix}}.
\label{init_cond_rho}
\end{equation}
In Fourier space, the initial condition for the conjugate density field translates to
\begin{equation}
    \tilde p_{\rho_1} (k,0)= \int_{-\infty}^{\infty}dx e^{-i k x} \theta (x)=-\frac{i}{k}+\pi \delta (k),
    \label{init_conjugate_rho}
\end{equation}
where $\delta (k)$ is the Dirac delta function. Using equations~\eqref{eigen_val},~\eqref{eigen_vec},~and \eqref{init_conjugate_rho}, we finally solve the matrix equation~\eqref{eq:matrix} to obtain
\begin{eqnarray}
\hspace{-0.5cm}
\tilde p_{\rho_1} (k,\tau) &=& e^{-\left(1+k^2\right) \tau}\tilde p_{\rho_1} (k,0) \left(\cosh \left(\tau \sqrt{1+k^2 g}\right)+\frac{\sinh \left(\tau
   \sqrt{1+k^2 g}\right)}{\sqrt{1-k^2 g^2}}\right),\label{p_rho_rho}\nonumber\\
\hspace{-0.5cm}
\tilde p_{m_1} (k,\tau) &=& \frac{i e^{-\left(1+k^2\right) \tau} \tilde p_{\rho_1} (k,0) ~k~ \text{Pe}(1-\bar{\rho}) \sinh \left(\tau \sqrt{1+k^2   g}\right)}{\sqrt{1+k^2 g}},
\label{p_rho}
\end{eqnarray}
where the expression for $\tilde p_{\rho_1} (k,0)$ is provided in equation~\eqref{init_conjugate_rho}.

We next compute the cumulants of the integrated current using the above exact expressions.
Using equations~\eqref{eq:rho0},~\eqref{first_cumulant1} and~\eqref{soln:zeroth_order_zero_mag}, the average integrated current can be directly obtained as
\begin{equation}
  \langle Q_{\rho}(T) \rangle_c= 0.
\label{first_cumulant_zero_mag}
\end{equation}
To compute the second cumulant of the integrated current,
we rewrite the expression for the second cumulant provided in equation~\eqref{second_cumulant2} as
\begin{eqnarray}
  \langle {Q_{\rho}(T)}^2 \rangle_c&=& \frac{1}{2\pi}  \int_{-\infty}^{\infty} dk \int_0^T dt \, \Bigg \{4 \rho_0 \tilde p_{m_1}(k,t)\tilde p_{m_1}(-k,t)\nonumber\\&&+
\sigma_{\rho_0}\tilde p_{\rho_1}(k,t)\tilde p_{\rho_1}(-k,t)k^2+
\sigma_{m_0}\tilde p_{m_1}(k,t)\tilde p_{m_1}(-k,t)k^2\Bigg \}.
\label{second_cumulant_zero_mag}
\end{eqnarray}
In the above equation we have used the fact that $\sigma_{\rho_0,m_0}=0$ as is clear from equations~\eqref{sigma} and~\eqref{soln:zeroth_order_zero_mag}. Equation~\eqref{second_cumulant_zero_mag} is a double integral and we can first compute the time integral. This yields the exact expression
\begin{eqnarray}
\langle {Q_{\rho}(T)}^2 \rangle_c&=& \frac{\sigma_{\bar{\rho}}}{8\pi}  \int_{-\infty}^{\infty} dk  \,  F(k,T),
\label{second_cumulantrho_zero_mag}
\end{eqnarray}
where $\sigma_{\bar{\rho}}$ is defined in equation~\eqref{sigma} and the function $F(k,T)$ is given as
\begin{equation}
   F(k,T)=e^{-2T h_1(k) } \left [ f_1(k,T)+f_2(k,T)+f_3(k,T)+f_4(k,T)\right ]. 
   \label{fkt}
\end{equation}
The function $F(k,t)$ is in turn composed of constituent functions $f_i$ (where $i=1,2,3,4$) and $h_1$. The exact expressions for these functions can be computed as
\begin{eqnarray}
f_1&=&\frac{2~ \text{Pe}^2  (1-\bar{\rho}) \left( h_1- k^2 \bar{\rho}
   \right)}{h_1 \left(1+g k^2\right)},\nonumber\\
   f_2&=&\frac{ e^{-2T \sqrt{1+g k^2} } \left(\sqrt{1+g
   k^2} - \text{Pe}^2 (1-\bar{\rho})(1+k^2 \bar{\rho})-1
   \right)}{h_1 \left(1+g
   k^2\right)+\left(1+g k^2\right)^{3/2}},\nonumber\\
   f_3&=&\frac{2~ e^{2 Th_1 }
   \left(\left(-2+g-k^2\right)
   h_1+\left(g+h_1
   \left(2-g+2 h_1\right)\right) \bar{\rho}
   \right)}{(2\bar{\rho}-1)\left( h_1^3-h_1
   \left(1+g k^2\right)\right)},\nonumber\\
   f_4&=&\frac{ e^{2 T\sqrt{1+g k^2} } \left(-\sqrt{1+g
   k^2} - \text{Pe}^2 (1-\bar{\rho})(1+k^2 \bar{\rho})-1\right)}{h_1
   \left(1+g k^2\right)-\left(1+g k^2\right)^{3/2}},\nonumber\\
   h_1&=&1+k^2~,
   \label{f_i(k,t)}
\end{eqnarray}
where the factor $g$ appearing in the above equations is defined in equation~\eqref{g}. 
The SSEP limit of the current fluctuations can be obtained by setting Pe$=0$ (which in turn corresponds to $g=0$) in equations~\eqref{second_cumulantrho_zero_mag}-\eqref{f_i(k,t)}. This sets $f_1=f_2=0$. We then obtain the simple result
\begin{eqnarray}
\langle {Q_{\rho}(T)}^2 \rangle_c&=&\frac{\sigma_{\bar{\rho}}}{4\pi}  \int_{-\infty}^{\infty} dk  \, e^{-2T h_1}\left(\frac{e^{2Th_1}-e^{2T}}{k^2}\right)=\sqrt{T} \frac{\sigma_{\bar{\rho}}}{\sqrt{2\pi}},~\text{for SSEP}.
\label{second_cumulantrho_ssep}
\end{eqnarray}
Therefore, for the SSEP, the fluctuations grow as $\sqrt{T}$ at all times. 

Interestingly, as we show below, the addition of activity introduces different scalings at different times in the current fluctuations. This can be shown through the computation of the variance of the integrated current in Laplace space. We also present an alternate method to derive the large and small time asymptotic behavior of the variance in appendix~\ref{appendix_asymptotics}. We define the Laplace transform of $\langle {Q_{\rho}(T)}^2 \rangle_c$ as $\mathcal{L}\left[ \langle {Q_{\rho}(T)}^2 \rangle_c\right]= \langle {\tilde Q_{\rho}(s)}^2 \rangle_c=\int_0^\infty dT~ \langle {Q_{\rho}(T)}^2 \rangle_c e^{-s T}$. Using equations~\eqref{second_cumulantrho_zero_mag}~and~\eqref{fkt}, we obtain the exact expression for the Laplace transform of $\langle {Q_{\rho}(T)}^2 \rangle_c$ as
\begin{small}
\begin{eqnarray}
\hspace{-2.2 cm}
\langle {\tilde Q_{\rho}(s)}^2 \rangle_c&=& \frac{\sigma_{\bar{\rho}}}{2\pi}  \int_{-\infty}^{\infty} dk  \,  \frac{(2+s) (4+s)+4 k^2 \left(k^2+3+s+\text{Pe}^2 (1-\bar{\rho}
   )^2\right)+4 \text{Pe}^2 (1-\bar{\rho} )}{s \left(2+2
   k^2+s\right) \left(4 k^4+s (4+s)+4 k^2
   (2+s-g)\right)}.
\label{second_cumulant_int_laplace}
\end{eqnarray}
\end{small}
After performing the $k$ integral in the above equation, we obtain the explicit expression for the variance in Laplace space. This expression is rather complicated, and we quote the asymptotic expansions in the small and large $s$ limits below. We have
\begin{small}
\begin{eqnarray}
\langle {\tilde Q_{\rho}(s)}^2 \rangle_c&\xrightarrow[s \rightarrow 0]{}& \frac{\sigma_{\bar{\rho}}~  \xi  \left(2+\text{Pe}^2 (1-\bar{\rho} )\right)}{4
   s^{3/2}}+\frac{\sigma_{\bar{\rho}} ^2~ \text{Pe}^2 \left(-1+g~ \xi ^3\right)}{8
   (1-g) s }+...~\text{for}~ g<2,
\label{rho_cums0}
\end{eqnarray}
\end{small}
and
\begin{eqnarray}
\langle {\tilde Q_{\rho}(s)}^2 \rangle_c&\xrightarrow[s \rightarrow \infty]{}&\frac{\sigma_{\bar{\rho}}}{2 \sqrt{2} s^{3/2}}+\frac{ \sigma_{\bar{\rho}} ^2~\text{Pe}^2}{16 \sqrt{2} s^{5/2}}+...~,
\label{rho_cumsinf}
\end{eqnarray}
where 
\begin{equation}
    \xi=\frac{1}{\sqrt{2-g}},
\end{equation}
is the correlation length derived in~\cite{agranov2021exact} and $g$ is defined in equation~\eqref{g}. We notice that $g < 2$ is the region where the homogeneous solutions $\rho_0(x,t)=\bar{\rho}$ and $m_0(x,t)=0$ to the noiseless hydrodynamic equations are linearly stable~\cite{kourbane2018exact}. For $g > 2$, which in turn corresponds to the linearly unstable region, the integral in equation~\eqref{second_cumulant_int_laplace} is not convergent. 

To obtain the behavior of the current fluctuations in the time domain, we invert the expressions provided in equation~\eqref{rho_cums0}~and~\eqref{rho_cumsinf}. This yields the large and small time behaviors of the variance as
\begin{small}
\begin{eqnarray}
\hspace{-1 cm}
\langle { Q_{\rho}(T)}^2 \rangle_c&\xrightarrow[T \rightarrow \infty]{}& \sqrt{T}\frac{\sigma_{\bar{\rho}}~  \xi  \left(2+\text{Pe}^2 (1-\bar{\rho} )\right)}{2
   \sqrt{\pi}}+\frac{\sigma_{\bar{\rho}} ^2~ \text{Pe}^2 \left(-1+g~ \xi ^3\right)}{8
   (1-g)  }+...~\text{for}~ g<2,
\label{rho_cumtinfty_exp}
\end{eqnarray}
\end{small}
and
\begin{eqnarray}
\langle { Q_{\rho}(T)}^2 \rangle_c&\xrightarrow[T \rightarrow 0]{}&\sqrt{T}\frac{\sigma_{\bar{\rho}}}{ \sqrt{2 \pi} }+T^{3/2}\frac{ \sigma_{\bar{\rho}} ^2~\text{Pe}^2}{12 \sqrt{2 \pi}}+...~.
\label{rho_cumt0_exp}
\end{eqnarray}
The leading term in the $T \rightarrow 0$ expansion of the variance corresponds to the current fluctuations in the SSEP starting from a quenched uniform density profile given in equation~\eqref{second_cumulantrho_ssep}. It is important to note that the above expressions are in the rescaled variables, and the unscaled expressions can be obtained from these using the substitution $\langle {Q_{\rho}(T)}^2 \rangle_c \xrightarrow{}\sqrt{D/\gamma}~\langle {Q_{\rho}(\gamma T)}^2 \rangle_c$.

Next, using the asymptotic behavior of the variance in equation~\eqref{rho_cumt0_exp}, we can extract the timescale $T^*$ up to which the system exhibits the short-time SSEP behavior. This can be obtained by equating the first two terms in the rhs yielding
\begin{equation}
  T^* \approx \frac{D}{\lambda^2 \sigma_{\bar{\rho}}}.
  \label{rho_0.5_t1}
\end{equation}
Interestingly, this timescale  diverges in the limit of zero density and therefore emerges purely from the interactions between particles.
At large times, a single RTP displays diffusive behavior beyond $ T \gg 1/\gamma$, with an effective diffusion constant $D_{\text{eff}}=D+\lambda^2/(2 \gamma)$~\cite{jose2022active,jose2022first,malakar2018steady}. We thus expect a $\sqrt{T}$ behavior of the current fluctuations beyond this timescale. We note, however, that the model studied in this paper does not trivially reduce to the SSEP with the dynamics governed by the single particle $D_\text{eff}$, as is clear from the large time asymptotic behavior of the current fluctuations in equation~\eqref{rho_cumtinfty_exp}.
 
\begin{figure} [!t]
 \includegraphics[width=1.0\linewidth]{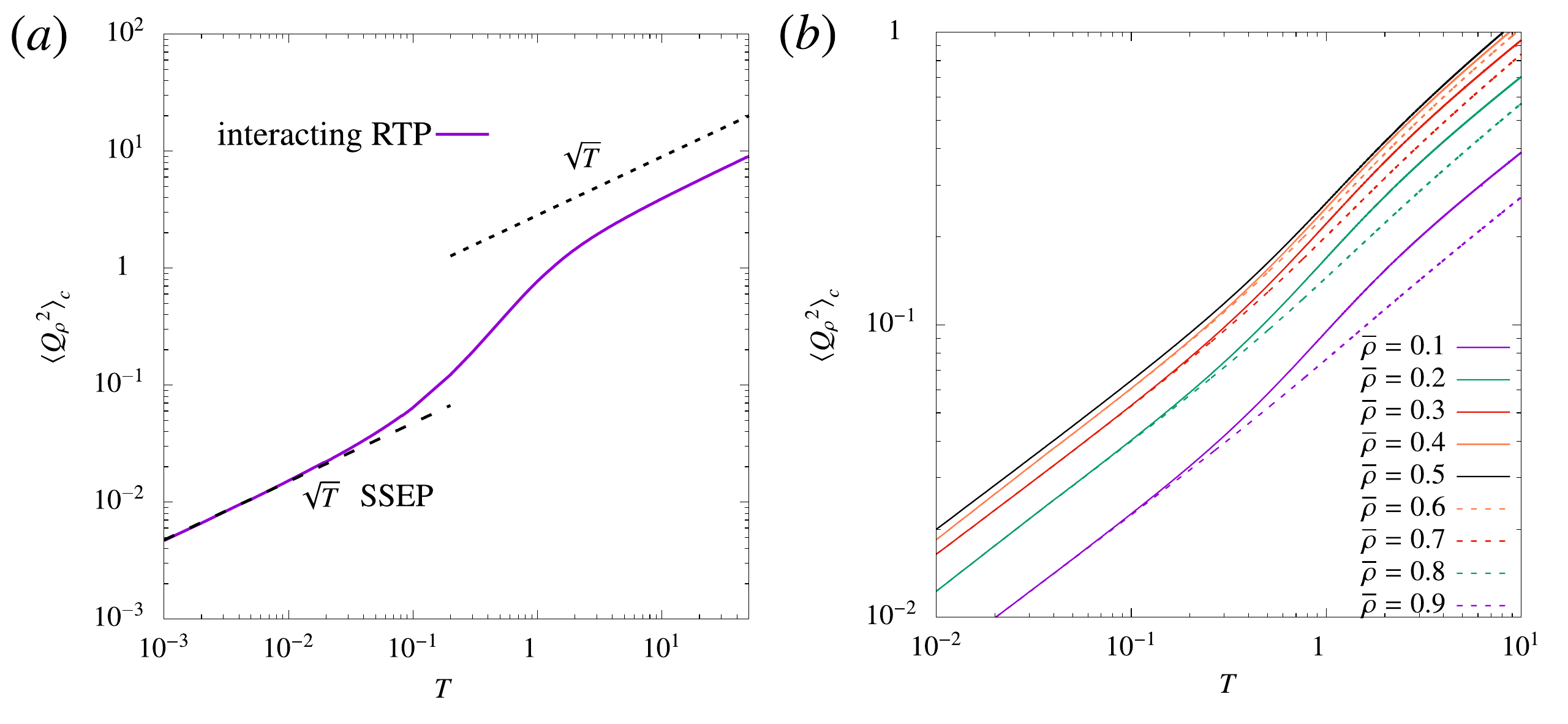}
\caption{
(a) The three regimes in the second cumulant of the integrated density current plotted as a function of time for the initial condition $\rho(x,0)=\bar{\rho}=0.25$ and $m(x,0)=0$. For this, we have used equation~\eqref{second_cumulantrho_zero_mag} and have done a numerical integration in Fourier space. The dashed curve at short times corresponds to the limiting case of SSEP. We observe that at small timescales $(T \rightarrow 0)$, the second cumulant reduces to that of a SSEP, and ${\langle {Q_{\rho}}^2 (T)\rangle}_c=\sqrt{T} \frac{\sqrt{D}\sigma_{\bar{\rho}}}{\sqrt{2\pi}}$ with $\sigma_{\bar{\rho}}=2 \bar{\rho}(1-\bar{\rho})$. At large time scales, the fluctuations again exhibit a $\sqrt{T}$ behavior. The spatial and temporal coordinates in this plot only have diffusive scaling. The different parameter values used are $D=1,~\gamma=1$ and~$\lambda=10$. (b) The second cumulant of the integrated density current plotted as a function of time for the initial condition $\rho(x,0)=\bar{\rho}$ and $m(x,0)=0$ for different values of $\bar{\rho}$. The Pe number is fixed to be $2$. We observe that the fluctuations are non-monotonic functions of the mean density. Initially, the fluctuations are symmetric about the density $0.5$ as in the SSEP. At large times, we observe that each of the solid curves splits into two and the system breaks particle-hole symmetry. The dashed curves correspond to the higher-density counterparts. The above plot is in the rescaled coordinates (diffusive scaling is always present).}
\label{three_regimeS_3ensity_dependence}
\end{figure}


We plot the second cumulant of the integrated density current as a function of time for the initial condition $\rho(x,0)=0.25$ and $m(x,0)=0$ in figure~\ref{three_regimeS_3ensity_dependence}(a). For this, we have used equation~\eqref{second_cumulantrho_zero_mag} and have done a numerical integration in Fourier space. The curve typically consists of three regimes; a short time $\sqrt{T}$ behavior predicted exactly by the SSEP, a cross-over regime, and a large time $\sqrt{T}$ behavior where activity drives large fluctuations. We also plot the second cumulant of the integrated density current as a function of time for the initial condition $\rho(x,0)=\bar{\rho}$ and $m(x,t)=0$ for different values of $\bar{\rho}$ with fixed Pe in figure~\ref{three_regimeS_3ensity_dependence}(b). We observe that the fluctuations are non-monotonic functions of the mean density. Initially, the fluctuations are symmetric about the density $0.5$ as in the case of the SSEP. That is, the fluctuations for densities $0.5+\Delta$ and $0.5-\Delta$ are exactly the same (where $0<\Delta <0.5$). At large times, each of the solid curves split into two. The dashed curves correspond to the higher density ($0.5+\Delta$) counterparts. At large times, the fluctuations for densities $0.5-\Delta$ are higher than for the densities $0.5+\Delta$ pointing to the lack of particle-hole symmetry in the model. However, at short times; where activity does not play any role, we recover the particle-hole symmetry associated with the symmetric exclusion process. 

\begin{figure} [!t]
 \includegraphics[width=1.0\linewidth]{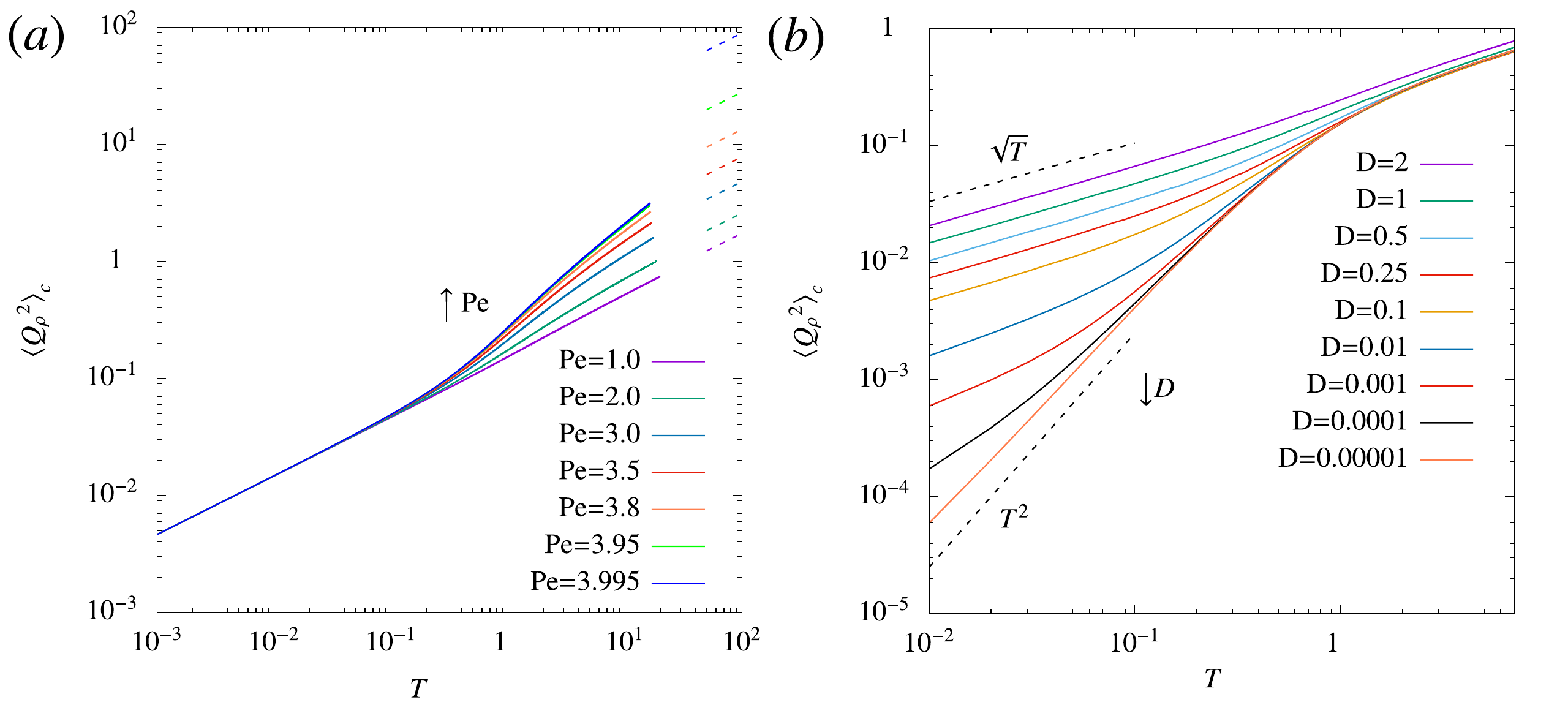}
\caption{(a)
The second cumulant of the integrated density current plotted as a function of time for the initial condition $\rho(x,0)=0.75$, $m(x,0)=0$ for different values of Pe.  For this, we have used equation~\eqref{second_cumulantrho_zero_mag} and have done a numerical integration in Fourier space. Since $(\bar{\rho}=0.75$, Pe$=4)$ corresponds to the critical point of the model, we observe large fluctuations as we cross the critical point. For small values of Pe, the theory predicts the typical fluctuations. For Pe $>4$, the system enters into the unstable (spinodal) region where homogenous phases are no longer stable. The dashed curves correspond to our predictions for the asymptotic behavior. The above plot is in the rescaled coordinates.
(b) The second cumulant of the integrated density current plotted as a function of time for the initial condition $\rho(x,0)=0.25$, $m(x,0)=0$  for different values of $D$. The fixed parameter values used are $\gamma=1,~\lambda=2$. As we reduce $D$, the small-time behavior changes from $\sqrt{T}$ to $T^2$. Since we display the effect of the diffusion constant on current fluctuations, this plot has just the diffusive rescaling.}
\label{pe_dif_dependence}
\end{figure}

In figure~\ref{pe_dif_dependence}(a), we display the behavior of current fluctuations for different choices of the Péclet number Pe. The density is fixed to be $\bar{\rho}=0.75$. Since $(\bar{\rho}=0.75$, Pe$=4)$ corresponds to the critical point~\cite{tailleur2008statistical} of the model, we observe large current fluctuations as we cross the critical point. For small values of Pe, the theory predicts the typical fluctuations. For Pe $>4$, the system enters into the unstable (spinodal) region where homogenous phases are no longer stable.  Another interesting feature we observe is that as we reduce the diffusion rate $D$ in the original active gas model keeping all other parameters fixed, the small time behavior of the current fluctuations gradually changes from $\sqrt{T}$ to $T^2$ in the $D \xrightarrow{} 0 $ limit. This behavior is clearly exhibited in figure~\ref{pe_dif_dependence}(b) for the average density $\bar{\rho}=0.25$. We discuss the zero diffusion limit in detail in section~\ref{sec:int_zero_dif}.
\subsubsection{Exact expression of the variance for \texorpdfstring{$\bar{\rho}=1/2$}{TEXT}}
For $\bar{\rho}=1/2$, which in turn corresponds to $g=0$ and $\xi={1}/{\sqrt{2}}$, the integral in~equation~\eqref{second_cumulantrho_zero_mag} can be computed exactly. This yields the exact expression,
\begin{eqnarray}
\langle {Q_{\rho}(T)}^2 \rangle_c&=& \frac{\sqrt{T}\left(4+\text{Pe}^2\right) }{8 \sqrt{2 \pi }}-\frac{e^{-2 T}
   \text{Pe}^2 \sinh ^2(T)}{16 \sqrt{2 \pi } \sqrt{T}}-\frac{1}{32} \text{Pe}^2
   \text{Erf}\left[\sqrt{2 T}\right],
\label{second_cumulantrho_zero_mag_rho05}
\end{eqnarray}
where Erf is the error function. We note that the above expression can be used to compute the behavior of fluctuations at all times. This expression indeed reproduces the limiting behaviors provided in equations~\eqref{rho_cumtinfty_exp} and~\eqref{rho_cumt0_exp} for $\bar{\rho}=1/2$. 

\subsubsection{Non-interacting limit}
To obtain the non-interacting limit of the current fluctuations for the active lattice model, we take a $\bar{\rho} \rightarrow 0$ limit of equation~\eqref{second_cumulant_int_laplace}. Performing a series expansion and keeping terms linear in $\bar{\rho}$ yields the expression for the Laplace transform of $\langle {Q_{\rho}(T)}^2 \rangle_c$ in the non-interacting limit. We have
\begin{small}
\begin{eqnarray}
\langle {\tilde Q_{\rho}(s)}^2 \rangle_c&=& \frac{ \bar{\rho}}{\pi}  \int_{-\infty}^{\infty} dk  \,  \frac{(2+s) (4+s)+4 k^2 \left(k^2+3+s+\text{Pe}^2 \right)+4 \text{Pe}^2 }{s \left(2+2
   k^2+s\right) \left(4 k^4+s (4+s)+4 k^2
   (2+s+\text{Pe}^2)\right)}.
\label{second_cumulant_nonint_laplace}
\end{eqnarray}
\end{small}
As for the interacting case, the $k$ integral in the above expression can be computed explicitly. Since this expression is long, we provide asymptotic expansions of the variance in Laplace space below. In the small $s$ limit we have
\begin{small}
\begin{eqnarray}
\langle {\tilde Q_{\rho}(s)}^2 \rangle_c&\xrightarrow[s \rightarrow 0]{}& \frac{\sqrt{2+\text{Pe}^2} \bar{\rho} }{2 s^{3/2}}-\frac{\text{Pe}^2 \left(12+5 \text{Pe}^2\right) \bar{\rho} }{16
   \left(2+\text{Pe}^2\right)^{3/2} \sqrt{s}}+...~,
\label{rho_cums0_non_int}
\end{eqnarray}
\end{small}
and in the large $s$ limit we have
\begin{eqnarray}
\langle {\tilde Q_{\rho}(s)}^2 \rangle_c&\xrightarrow[s \rightarrow \infty]{}&\frac{\bar{\rho} }{\sqrt{2} s^{3/2}}+\frac{5 \text{Pe}^2 \bar{\rho} }{4
   \sqrt{2} s^{7/2}}+...~.
\label{rho_cumsinf_non_int}
\end{eqnarray}
Inverting the above expressions yield the large and small time behaviors of the variance in the non-interacting limit as
\begin{small}
\begin{eqnarray}
\langle { Q_{\rho}(T)}^2 \rangle_c&\xrightarrow[T \rightarrow \infty]{}& \sqrt{T}\frac{\sqrt{2+\text{Pe}^2}~  \bar{\rho} }{\sqrt{\pi }}-\frac{1}{\sqrt{T}}\frac{\text{Pe}^2 \left(12+5 \text{Pe}^2\right) \bar{\rho} }{16
   \left(2+\text{Pe}^2\right)^{3/2} \sqrt{\pi } }+...~,
\label{rho_cumtinfty_exp_nonint}
\end{eqnarray}
\end{small}
and
\begin{eqnarray}
\langle { Q_{\rho}(T)}^2 \rangle_c&\xrightarrow[T \rightarrow 0]{}&\sqrt{T}\frac{\sqrt{2} \bar{\rho}}{ \sqrt{ \pi} }+T^{5/2}\frac{ \sqrt{2} \bar{\rho} ~\text{Pe}^2}{3 \sqrt{ \pi}}+...~.
\label{rho_cumt0_exp_nonint}
\end{eqnarray}
We note that the subleading corrections at large and short times are of order $T^{-1/2}$ and  $T^{5/2}$ respectively. This is in contrast to the subleading behaviors provided in equations~\eqref{rho_cumtinfty_exp}~and~\eqref{rho_cumt0_exp}, where the corrections are of order $T^{0}$ and  $T^{3/2}$ respectively. This is due to the fact that the coefficients of these subleading terms in equations~\eqref{rho_cumtinfty_exp}~and~\eqref{rho_cumt0_exp} do not have terms linear in $\bar{\rho}$, and therefore vanish in the $\bar{\rho} \to 0$ limit.

In the unscaled coordinates, the leading order terms in the asymptotic limits have the explicit forms
\begin{eqnarray}
\langle {Q_{\rho}(T)}^2 \rangle_c&\xrightarrow[T \rightarrow \infty]{}& \sqrt{T} \frac{\sqrt{D_{\text{eff}}}}{\sqrt{2\pi}}2\bar{\rho},
\label{rho_cumtinf_non_int}
\end{eqnarray}
and
\begin{eqnarray}
\langle {Q_{\rho}(T)}^2 \rangle_c&\xrightarrow[T \rightarrow 0]{}& \sqrt{T} \frac{\sqrt{D}}{\sqrt{2\pi}}2\bar{\rho},
\label{rho_cumt0_non_int}
\end{eqnarray}
where $D_{\text{eff}}=D+\lambda^2/(2 \gamma)$ is the effective diffusion constant for a single RTP with diffusion in one dimension. 

The factor $2 \bar{\rho}$ appearing in equations \eqref{rho_cumtinf_non_int}~and~\eqref{rho_cumt0_non_int} arises because we consider uniform initial conditions, with particles initially distributed uniformly on both sides of the origin. For instance, if we consider step initial conditions with particles uniformly distributed towards the left of the origin, this factor would be $\bar{\rho}$. As anticipated, we recover the $\sqrt{T}$ behavior of the current fluctuations described by the effective diffusion constant of the single RTP at large times. At short times, the particles behave as non-interacting random walkers, and the fluctuations display a $\sqrt{T}$ behavior described in equation~\eqref{rho_cumt0_non_int}, which is consistent with the expression for current fluctuations of non-interacting random walkers derived in~\cite{derrida2009current2}. The typical timescale $T^*$ up to which the current fluctuations exhibit the short time $\sqrt{T}$ behavior can be computed by equating the first two terms in the rhs of equation~\eqref{rho_cumt0_exp_nonint}. This yields
\begin{equation}
    T^*\approx \frac{\sqrt{D}}{\sqrt{\gamma}\lambda}.
    \label{time_scale_non_int}
\end{equation}
This timescale can be understood as follows. Since we focus on the low-density limit, the current fluctuations at short times are dominated by single particle fluctuations. For quenched initial conditions, each particle can be considered as initialized in either of the bias states $+$ or $-$. For a single particle initialized asymmetrically in the $+$ or $-$ state, the mean squared displacement behaves as
\begin{eqnarray}
{\langle x^2 \rangle}_c&\xrightarrow[T \rightarrow 0]{}&2DT+\frac{4}{3}\gamma\lambda^2 T^3+...~.    
\end{eqnarray}
The typical timescale up to which the system exhibits the short-time diffusive behavior is thus obtained as $T^* \approx {\sqrt{  D}}/{(\sqrt{\gamma}\lambda)} $, consistent with the timescale derived in equation~\eqref{time_scale_non_int}.


\subsection{Interacting active lattice gas in the zero diffusion limit 
}
\label{sec:int_zero_dif}
We next derive asymptotic limits of the variance of the integrated current in the active lattice gas model with zero diffusion which can be analyzed as a limiting case of the model studied in this paper. The $D \rightarrow 0$ limit allows us to understand the limiting behavior observed in the current fluctuations as the microscopic diffusion constant is reduced to a very small value. The hydrodynamic equations for the $\rho$ and $m$ fields are valid for any small non-zero $D$. In this limit, we expect the hydrodynamic scaling to still be valid, and the diffusion term in the hydrodynamic equations can be neglected. 
The current fluctuations in the $D \rightarrow 0$ limit are illustrated in figure~\ref{pe_dif_dependence}(b). When the diffusion constant is decreased, a regime where the fluctuations grow as $T^2$ begins to appear. 
To further characterize this, we study the MFT equations with $D = 0$, which allows us to derive analytic expressions for the observed limiting $T^2$ behavior as the diffusion constant is reduced.
In the zero diffusion limit, the fluctuating hydrodynamic equations provided in equation~\eqref{eq:fluctuating_hydro} reduce to
\begin{eqnarray}
\partial_t \rho&=& -\lambda\partial_x[m(1-\rho)],\nonumber\\
\partial_t m&=& -\lambda\partial_x[\rho(1-\rho)]-2\gamma m+\frac{2}{\sqrt{L}} \sqrt{\gamma}\,\eta_K . \label{eq:m_unscaled_zero_diff}
\end{eqnarray}
As in the diffusive case, we analyze the case of uniform initial conditions which is provided in equation~\eqref{init_fields}. The zeroth order equations provided in equation~(\ref{zeroth_order_eqns}) admit the analytical solutions in equation~\eqref{soln:zeroth_order_zero_mag}.
Substituting these solutions into the first-order equations for the conjugate fields yields
\begin{eqnarray}
  \partial_t {p_\rho}_1   &=&  - \lambda (1 - 2\bar{\rho}) \partial_x {p_m}_1,\nonumber\\
 \partial_t {p_m}_1 &=&    - \lambda  (1 - \bar{\rho}) \partial_x {p_\rho}_1+2 \gamma{p_m}_1.
\label{eq:zero_mag_zero_dif}
\end{eqnarray}
Notice that the above equations are in the unscaled coordinates and have just the diffusive rescaling.

Taking a Fourier transform of equation~(\ref{eq:zero_mag_zero_dif}) yields the matrix equation
\begin{equation}
\frac{\partial}{\partial \tau}
\ket{\tilde p (k,\tau)}
=\mathcal{M}_0(k)
\ket{\tilde p (k,\tau)},
\label{eq:matrix_zero_dif}
\end{equation}
where the column vector $\ket{\tilde p (k,\tau)}$ is defined in equation~\eqref{column_vec} 
and
\begin{equation}
    \mathcal{M}_0(k)=\begin{pmatrix}
 0 & -i k\lambda(2\bar{\rho}-1) \\
i k \lambda(1-\bar{\rho}) & -2\gamma
 \end{pmatrix}.
\end{equation}
Equation~(\ref{eq:matrix_zero_dif}) can be solved by diagonalizing the matrix $\mathcal{M}_0(k)$. 
The eigenvalues $\zeta_1(k)$, $\zeta_2(k)$ and the eigenvectors $\ket{\phi_1(k)}$, $\ket{\phi_2(k)}$ of the matrix $\mathcal{M}_0(k)$ are given as
\begin{equation}
\zeta_1(k)=-\gamma-\sqrt{\gamma^2+k^2 \lambda^2 g_0},~ \zeta_2(k)=-\gamma+\sqrt{\gamma^2+k^2 \lambda^2 g_0}~,  
\label{eigen_val_zero_diff}
\end{equation}
and
\begin{equation}
\ket{\phi_1(k)}=    {\begin{pmatrix}
\frac{i \left(-\gamma+\sqrt{\gamma^2+k^2 \lambda^2 g_0}\right)}{k \lambda (1-\bar{\rho})}\\
 1
\end{pmatrix}},~\ket{\phi_2(k)}=    {\begin{pmatrix}
\frac{i \left(-\gamma-\sqrt{\gamma^2+k^2 \lambda^2 g_0}\right)}{k \lambda (1-\bar{\rho})}\\
 1
 \label{eigen_vec_zero_diff}
\end{pmatrix}}.
\end{equation}
The constant $g_0$ appearing in the above equation has the explicit form
\begin{equation}
    g_0=(1-\bar{\rho})(2 \bar{\rho}-1).
\label{g0}
\end{equation}
We use the initial condition in equation~\eqref{init_cond_rho} along the expressions for the eigenvalues and the eigenvectors in equations~\eqref{eigen_val_zero_diff} and~\eqref{eigen_vec_zero_diff} to solve the matrix equation in~\eqref{eq:matrix_zero_dif}.
The final expressions for the conjugate fields are thus obtained as
\begin{eqnarray}
\hspace{-1.5cm}
\tilde p_{\rho_1} (k,\tau) =e^{-\gamma \tau}\tilde p_{\rho_1} (k,0) \left(\cosh \left(\tau \sqrt{\gamma^2+k^2 \lambda^2 g_0}\right)+\frac{\gamma \sinh \left(\tau
   \sqrt{\gamma^2+k^2 \lambda^2 g_0}\right)}{\sqrt{\gamma^2+k^2 \lambda^2 g_0}}\right),\label{p_rho_rho_zero_diff}\nonumber\\
\hspace{-1.5cm}
\tilde p_{m_1} (k,\tau) = \frac{i e^{-\gamma \tau} \tilde p_{\rho_1} (k,0) ~k~ \lambda(1-\bar{\rho}) \sinh \left(\tau \sqrt{\gamma^2+k^2 \lambda^2 g_0}\right)}{\sqrt{\gamma^2+k^2 \lambda^2 g_0}}.
\label{p_rho_zero_diff}
\end{eqnarray}

We next compute the cumulants associated with the integrated current. As in the diffusive case, the mean integrated current, $\langle {Q_{\rho}(T)} \rangle_c$ is zero. Using equation~\eqref{second_cumulant2} the second cumulant of the integrated density current assumes the form
\begin{equation}
  \langle {Q_{\rho}(T)}^2 \rangle_c=   \int_{-\infty}^{\infty} dx \int_0^T dt \, 4 \gamma \rho_0 {{p_m}_1}^2.
\label{second_cumulant3}
\end{equation}
The other two terms in equation~\eqref{second_cumulant2} are zero due to zero diffusion.
In Fourier space, the above equation can be rewritten as
\begin{eqnarray}
  \langle {Q_{\rho}(T)}^2 \rangle_c&=& \frac{1}{2\pi}  \int_{-\infty}^{\infty} dk \int_0^T dt \, 4 \gamma \rho_0 \tilde p_{m_1}(k,t)\tilde p_{m_1}(-k,t).
\label{second_cumulant_zero_mag_zero_dif}
\end{eqnarray}
The time integral in equation~\eqref{second_cumulant_zero_mag_zero_dif} can be first computed explicitly. Using equations~\eqref{p_rho_zero_diff} and~\eqref{second_cumulant_zero_mag_zero_dif}, we thus obtain
\begin{eqnarray}
  \langle {Q_{\rho}(T)}^2 \rangle_c&=& \frac{1}{2\pi} \frac{\gamma}{\lambda} \int_{-\infty}^{\infty} d\alpha~ G(\alpha,T),
\label{second_cumulant_zero_mag_zero_dif_exp_2}
\end{eqnarray}
where
\begin{eqnarray}
G(\alpha,T)&=&  \frac{e^{-2 T \gamma } \lambda ^2 
   (1-\bar{\rho}) \bar{\rho}}{\alpha ^2 \gamma ^2 \left(1+g_0 \alpha
   ^2\right)  (2 \bar{\rho}-1)} \times \Big [\sqrt{1+g_0 \alpha ^2} \sinh
   \left(2 T \gamma  \sqrt{1+g_0\alpha
   ^2}\right) \nonumber\\&&+ \cosh \left(2 T
   \gamma  \sqrt{1+g_0 \alpha
   ^2}\right) +g_0 \alpha ^2-e^{2 T \gamma } \left(1+g_0
   \alpha ^2\right)\Big].
\label{Gkt}
\end{eqnarray}
In the above expression, we have used the substitution, $k=\alpha \gamma/\lambda$.

We next define the Laplace transform of the function $G(\alpha,T)$ as $\mathcal{L}\left[ G(\alpha,T)\right]=\tilde G(\alpha,s)=\int_0^\infty dT~ G(\alpha,T)e^{-s T}$. Using equation~\eqref{Gkt}, we obtain the exact expression for the Laplace transform of $G(\alpha,T)$ as
\begin{eqnarray}
\tilde G(\alpha,s)&=&  \frac{8 g_0 \gamma  \lambda ^2
   (1-\bar{\rho} ) \bar{\rho} }{s (s+2
   \gamma ) \left(s^2+4 s
   \gamma -4 g_0 \alpha ^2
   \gamma ^2\right) (2 \bar{\rho}
   -1)}.
\label{Gks}
\end{eqnarray}
Integrating the above function over $\alpha$ yields
\begin{eqnarray}
\int_{-\infty}^{\infty} d\alpha~ \tilde G(\alpha,s)=\frac{4 \sqrt{|g_0|} \pi  \lambda ^2
   (1-\bar{\rho} ) \bar{\rho} }{s (s+2
   \gamma ) \sqrt{ s (s+4
   \gamma )} (1-2 \bar{\rho} )},~ g_0 < 0,
\label{Gs}
\end{eqnarray}
where $g_0$ is defined in equation~\eqref{g0}.
Using equation~\eqref{second_cumulant_zero_mag_zero_dif_exp_2}, we now obtain
\begin{eqnarray}
  \langle {Q_{\rho}(T)}^2 \rangle_c&=& \frac{1}{2\pi} \frac{\gamma}{\lambda} \mathcal{L}^{-1}\left[\int_{-\infty}^{\infty} d\alpha~ \tilde G(\alpha,s)\right].
\end{eqnarray}
We can compute this inverse Laplace transform yielding the final expression for the variance as
\begin{eqnarray}
\langle {Q_{\rho}(T)}^2 \rangle_c&=& \frac{\sqrt{|g_0|} (1-\bar{\rho} )\bar{\rho}\lambda T  e^{-2 \gamma  T}}{2(1-2\bar{\rho} )} \times\nonumber\\&&  \Big[(2+\pi  \pmb{L}_0(2 T \gamma )) \pmb{I}_1(2 T
 \gamma )-\pi \pmb{L}_1(2 T \gamma ) \pmb{I}_0(2 T \gamma )\Big],~ g_0 < 0.
\label{second_cumulantrho_zero_mag_final_exp}
\end{eqnarray}
Here, $\pmb{L}_0$ and $\pmb{L}_1$ are modified Struve functions and $\pmb{I}_0$ and $\pmb{I}_1$ are modified Bessel functions.
The above expression provides the full time dependence of the variance.
Finally, we compute the leading order terms in the asymptotic expansions of the variance as
\begin{eqnarray}
\hspace{-0.3cm}
\langle {Q_{\rho}(T)}^2 \rangle_c&\xrightarrow[T \rightarrow \infty]{}&\sqrt{T}\frac{ \sqrt{|g_0|} \lambda  (1-\bar{\rho} ) \bar{\rho}
   }{\sqrt{\pi } \sqrt{\gamma } (1-2 \bar{\rho}
   )},
\label{second_cumulantrho_zero_mag_zero_dif_tinfty}
\end{eqnarray}
and
\begin{eqnarray}
\hspace{-0.3cm}
\langle {Q_{\rho}(T)}^2 \rangle_c&\xrightarrow[T \rightarrow 0]{}&{T}^2\frac{ \sqrt{|g_0|}\gamma \lambda  (1-\bar{\rho} ) \bar{\rho}
   }{ (1-2 \bar{\rho}
   )}.
\label{second_cumulantrho_zero_mag_zero_dif_t0}
\end{eqnarray}
In the above expressions, $g_0 < 0$ defines the region where the homogeneous solutions $\rho(x,t)=\bar{\rho}$ and $m(x,t)=0$ to the hydrodynamic equations with $D=0$ are linearly stable. This is equivalent to the limit $\bar{\rho}<0.5$. For densities $\ge 0.5$, the integral in equation~\eqref{Gs} is not convergent. For the case with a finite diffusive constant, we were able to derive the exact expression for the second cumulant at all times for density~$\bar{\rho}=0.5$. However, for the case with zero diffusion, $\bar{\rho}=0.5$ is the critical point of the model, and the hydrodynamic description fails. For $T\gg1/\gamma$, the system becomes effectively diffusive and the current fluctuations exhibit a $\sqrt{T}$ behavior. This can also be seen by equating the expressions in equations~\eqref{second_cumulantrho_zero_mag_zero_dif_tinfty} and~\eqref{second_cumulantrho_zero_mag_zero_dif_t0}. Therefore, for timescales larger than $1/\gamma$, the system becomes effectively diffusive.
\begin{figure} [!t]
 \includegraphics[width=1.0\linewidth]{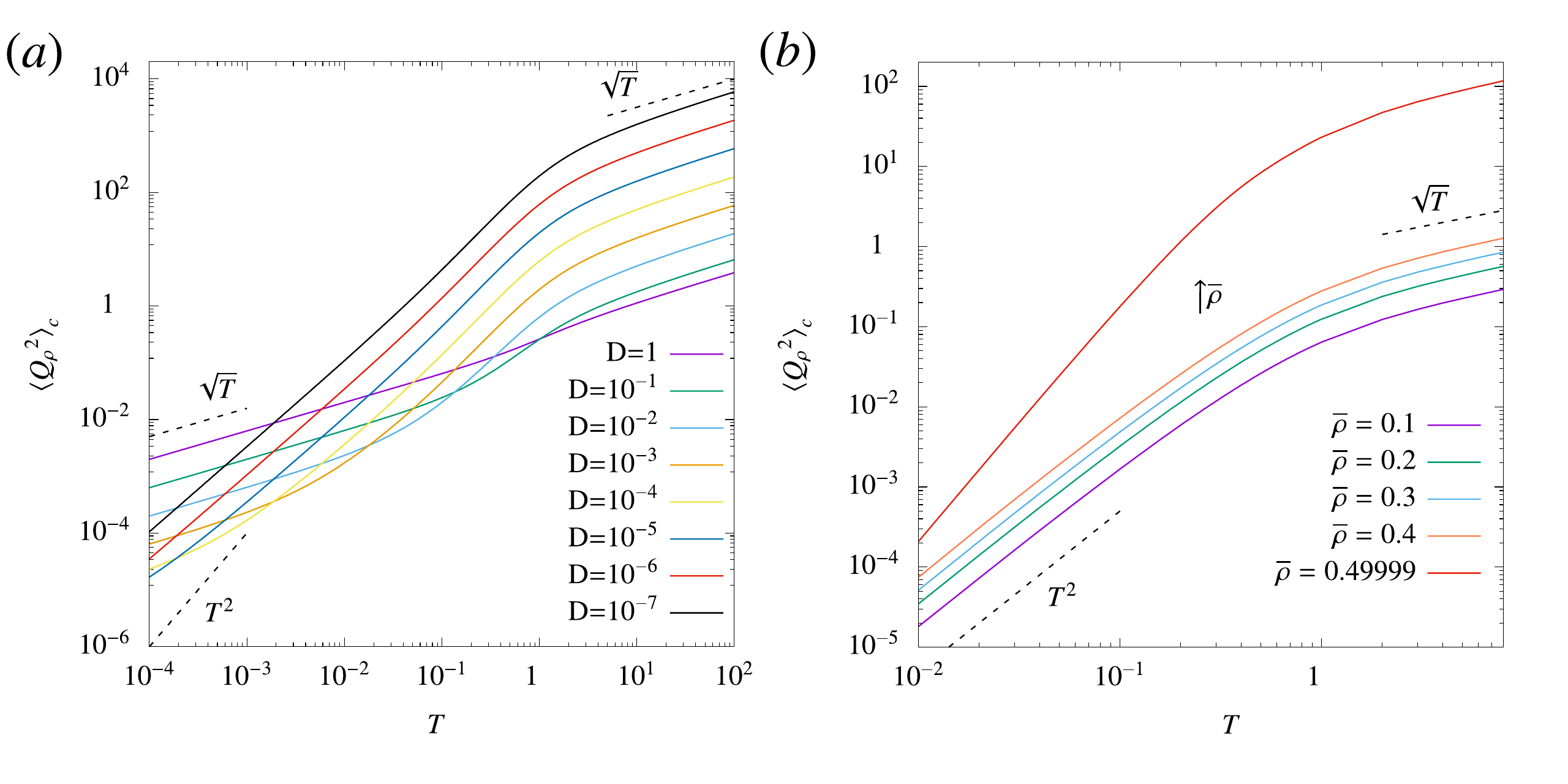}
\caption{(a)
The second cumulant of the integrated density current plotted as a function of time for the initial condition $\rho(x,0)=0.5$, $m(x,0)=0$ for different values of the diffusion constant. For this, we have used the equation~\eqref{second_cumulantrho_zero_mag} and have done a numerical integration in Fourier space. Since $\bar{\rho}=0.5$,  corresponds to the critical point of the active lattice gas model with zero diffusion, we observe large fluctuations as we approach the critical point. The fixed parameter values used are $\gamma=1,~\lambda=2$. The above plot has only the diffusive scaling.
(b) The second cumulant of the integrated density current for the zero diffusive active system provided in equation~\eqref{second_cumulantrho_zero_mag_final_exp} plotted as a function of time for the initial condition $\rho(x,0)=0.5$, $m(x,0)=0$  for different values of $\bar{\rho}$. The fixed parameter values used are $D=0,~\gamma=1,~\lambda=2$. As we approach the density $\bar{\rho}=0.5$, the system exhibits large current fluctuations. This plot has just the diffusive rescaling.}
\label{dens_5}
\end{figure}

We have plotted the second cumulant for different diffusion constants with average density $\bar{\rho}=0.5$ in figure~\ref{dens_5}(a). The behavior of the fluctuations is very different for the two choices of densities as is evident from figures~\ref{pe_dif_dependence}(b) and~\ref{dens_5}(a). Since $\bar{\rho}=0.5$,  corresponds to the critical point of the active lattice gas model with zero diffusion, we observe large fluctuations as $D \rightarrow 0$. In figure~\ref{dens_5}(b), we have also plotted the current fluctuations for the zero diffusive active gas model for different densities. The current fluctuations diverge as we approach the critical point $\bar{\rho}=0.5$.
\subsubsection{Non-interacting limit}
To obtain the non-interacting limit, it is sufficient to take a $\bar{\rho} \xrightarrow{} 0$ limit of the expression provided in equation~\eqref{second_cumulantrho_zero_mag_final_exp}. This yields
\begin{eqnarray}
\hspace{-1 cm}
\langle {Q_{\rho}(T)}^2 \rangle_c&=& \frac{\bar{\rho}\lambda T  e^{-2 \gamma  T}}{2}  \Big[(2+\pi  \pmb{L}_0(2 T \gamma )) \pmb{I}_1(2 T
 \gamma )-\pi \pmb{L}_1(2 T \gamma ) \pmb{I}_0(2 T \gamma )\Big],
\label{second_cumulantrho_zero_mag_non_int_final_exp}
\end{eqnarray}
with the limiting behaviors
\begin{eqnarray}
\langle {Q_{\rho}(T)}^2 \rangle_c&\xrightarrow[T \rightarrow \infty]{}&\sqrt{T}\frac{\sqrt{D^0_{\text{eff}}} }{ \sqrt{2 \pi }} 2\bar{\rho},
\label{second_cumulantrho_zero_mag_zero_dif_tinfty_non_interacting}
\end{eqnarray}
and
\begin{eqnarray}
\langle {Q_{\rho}(T)}^2 \rangle_c&\xrightarrow[T \rightarrow 0]{}&{T}^2{ \gamma \frac{\lambda}{2}   (2\bar{\rho})
   }.
\label{second_cumulantrho_zero_mag_zero_dif_t0_non_interacting}
\end{eqnarray}
Here, $ D^0_{\text{eff}}= { \lambda^2}/{(2\gamma)}$ is the effective diffusion constant for a single non-diffusive RTP in one dimension~\cite{jose2022active,jose2022first}. The superscript ``$0$" indicates that diffusion is absent. For non-interacting RTPs without diffusion, it can also be shown using Green's function techniques that the variance displays the exact same behavior predicted in equation~\eqref{second_cumulantrho_zero_mag_non_int_final_exp} for quenched density and quenched magnetization initial conditions~\cite{jose2023generalized}. The large time behavior of the current fluctuations quoted in equation~\eqref{second_cumulantrho_zero_mag_zero_dif_tinfty_non_interacting} has also been derived in a continuous space non-interacting RTP model with a quenched initial condition for the density profile and an annealed initial condition for the magnetization profile~\cite{banerjee2020current}. 
Our result in equation~\eqref{second_cumulantrho_zero_mag_zero_dif_tinfty_non_interacting} is larger by a factor of $2$ as we consider a uniform profile, as opposed to a step initial condition~\cite{banerjee2020current}. 
It can also be seen from equations~\eqref{second_cumulantrho_zero_mag_zero_dif_t0} and~\eqref{second_cumulantrho_zero_mag_zero_dif_tinfty_non_interacting} that for quenched initial conditions, the $T^2$ behavior at short times holds for all densities, and not just in the low-density limit. 
For $T\gg1/\gamma$, the system becomes effectively diffusive with a modified diffusion constant $D^0_{\text{eff}}$. However, this effective diffusion constant does not appear when interactions are considered as is clear from equation~\eqref{second_cumulantrho_zero_mag_zero_dif_tinfty}. 

Although the above analytical results are obtained for the case of an infinite lattice, microscopic simulations done on a finite lattice provide another route to understanding the boundary effects on the fluctuations and verifying the small time asymptotics predicted from the infinite lattice
calculations. We provide details of such numerical simulations in appendix~\ref{appendix_simulations}.



\section{Conclusion and discussion}
\label{sec:conclusion}
In this paper, we have studied the current fluctuations in an interacting active lattice gas that allows a comparison between microscopic measurements and predictions based on fluctuating hydrodynamics and macroscopic fluctuation theory. We used this model and the associated macroscopic fluctuation theory to compute the cumulants of the time-integrated current through the origin. This was possible through an application of a perturbative approach to the Euler-Lagrange equations associated with the action appearing in the generating function of the integrated current. However, as the non-linear equations are hard to analyze in an exact manner, we used a simple initial condition with constant density and zero magnetization profiles. We found a very good match between the theoretical predictions and simulations of the microscopic dynamics of the model, further confirming the validity of the fluctuating hydrodynamic framework for this model. 

Interestingly, we found that the fluctuations of the integrated density current in an interacting active lattice gas display three regimes; (1) the first regime where the fluctuations are exactly given by the $\sqrt{T}$ behavior of the SSEP as shown in equation~\eqref{rho_cumt0_exp} (2) a cross-over regime where activity and interactions drive larger fluctuations (3) a third regime where the fluctuations again grow as $\sqrt{T}$, but with a coefficient that depends on the Péclet number along with the density and the initial arrangement of particles as in equation~\eqref{rho_cumtinfty_exp}. 
The two diffusive regimes originate from the short-time diffusive motion arising from the intrinsic rate $D$ and the late-time effective diffusion of the particles respectively. 
The first regime appears up to a typical timescale $T^* \approx { D/}{\lambda^2 \sigma_{\bar{\rho}}}$.
At late times, the motion of a single RTP becomes effectively diffusive for $T \gg 1/\gamma$. Therefore the $\sqrt{T}$ behavior at large times originates from this effective late-time diffusive behavior of RTPs.
In the non-interacting limit, the integrated current can be expressed as a function of the effective diffusion constant $D_{\text{eff}}$ as demonstrated in equation~\eqref{rho_cumtinf_non_int}.
However, the general interacting model does not reduce to the SSEP with dynamics governed by the single particle $D_{\text{eff}}$ at large times. In this case, the interplay of activity and interactions remains important, as evidenced by the prefactor of the leading order term in equation~\eqref{rho_cumtinfty_exp}.

For density $\bar{\rho}=1/2$, we were able to derive the exact expression for the fluctuations at all times for the interacting active particle model which is given in equation~\eqref{second_cumulantrho_zero_mag_rho05}. Additionally, in the limit of zero diffusion, we showed that the current fluctuations typically consist of two regimes with an initial $T^2$ behavior as in equation~\eqref{second_cumulantrho_zero_mag_zero_dif_t0} and a later $\sqrt{T}$ behavior as in equation~\eqref{second_cumulantrho_zero_mag_zero_dif_tinfty}. We also computed the full-time dependence of the fluctuations for the zero diffusive model which is provided in equation~\eqref{second_cumulantrho_zero_mag_final_exp}. The $T^2$ behavior of fluctuations at short times can be attributed to the ballistic motion of particles as well as the quenched initial conditions in the density and the magnetization fields~\cite{jose2023generalized}. Furthermore, in the non-interacting limit ($\bar{\rho} \rightarrow 0$), the fluctuations exhibit a large time $\sqrt{T}$ behavior which can be expressed in terms of the single particle effective diffusion constant $ D^0_{\text{eff}}$ as shown in equation~\eqref{second_cumulantrho_zero_mag_zero_dif_tinfty_non_interacting}. For higher densities, our results demonstrate that the late-time diffusive behavior of the current fluctuations is modified in a non-trivial manner due to interactions between particles, as evidenced by the coefficient of $\sqrt{T}$ in equation~\eqref{second_cumulantrho_zero_mag_zero_dif_tinfty}.

The model studied in this paper has within it, hydrodynamic instabilities, which have been shown to be the analogs of MIPS. In this case, the MFT approach is able to capture the non-trivial entropy production \cite{agranov2022entropy}, however, the integrated current does not seem to be amenable to these methods beyond the phase boundary. It would be interesting to analyze how the integrated current behaves beyond the phase boundary. Several interesting directions remain to be pursued. It would certainly be useful to study other initial conditions where the coupled non-linear equations representing the zeroth order solutions can be solved analytically. This would greatly simplify the analysis of the current fluctuations, as has been shown in the case of the SSEP. Since in this work, we have focused on the case with fixed initial conditions, it would be interesting to study the effects of activity on the differences between quenched and annealed settings in detail. Another interesting quantity to investigate is the integrated magnetization current which measures the excess of $+$ particles crossing the origin up to time $T$. The framework derived here  can also be used to study other models where multiple coupled fields can lead to phase separation and aggregation, such as the Light-Heavy model~\cite{lahiri1997steadily,chakraborty2019dynamics,mahapatra2020light}.

\section{Acknowledgments}
We thank Mustansir Barma for several useful suggestions at various stages of this work. We thank Prasad Perlekar, Alberto Rosso, Yariv Kafri, Sanjib Sabhapandit, Kirone Mallick, Paul Krapivsky, Abhishek Dhar, Dipanjan Mandal, Roshan Maharana, and Surajit Chakraborty for useful discussions. We are also grateful to the anonymous referees for their valuable comments and inputs. This project was funded by intramural funds at TIFR, Hyderabad from the Department of Atomic Energy (DAE).
\section*{Appendix}
\appendix
\section{Alternate expressions for the cumulants}
\label{appendix_alternate_exp}
The cumulants of the integrated current can also be computed using the following alternate method~\cite{dandekar2022macroscopic}. The scaled cumulant generating function (SCGF) of the integrated current $Q_{\rho}(T)$ across the origin up to time $T$ can be computed as 
\begin{equation}
\psi_{\rho}(\Lambda,T) = \log{\big\langle e^{\Lambda Q_{\rho}(T)} \big\rangle}, 
 \label{cgf}
\end{equation}
where the average is given by the ensemble weighted by the action $\cal{S}$. Differentiating equation~(\ref{cgf}) with respect to $\Lambda$ yields
\begin{equation}
\psi_{\rho}'(\Lambda,T) = \frac{\langle Q_{\rho} e^{\Lambda Q_{\rho}}\rangle}{\langle e^{\Lambda Q_{\rho}} \rangle} = \langle Q_{\rho} \rangle_{\Lambda}={Q_{\rho}}_{MFT},
\end{equation}
where the average $\langle Q_{\rho} \rangle_{\Lambda}$ is over the ensemble with the modified weight $S + \Lambda Q_{\rho}$. This gives the average current for a given $\Lambda$. This is the same as the MFT solution as the MFT equations provide the saddle-point solution to the modified action. The MFT solutions can be perturbatively expanded as
\begin{equation}
 {Q_{\rho}}_{MFT}=  {Q_{\rho_0}}_{MFT}+\Lambda {Q_{\rho_1}}_{MFT}.
\end{equation}
Using this expansion in equation~(\ref{psi_definition}), we directly obtain the expressions for the first and second cumulants 
provided in equations~\eqref{first_cumulant1} and \eqref{second_cumulant1}.
\section{Asymptotic expansions of the variance}
\label{appendix_asymptotics}
We next present an alternate method to compute the limiting behaviors of the variance of the integrated current for an interacting active lattice gas.
We use the substitution $u=k \sqrt{T}$ in equation~\eqref{second_cumulantrho_zero_mag} to extract the scaling behavior of the integrand $F(k,T)$ at small and large times. It is easy to show that $F(k,T)$ admits the scaling forms
\begin{eqnarray}
    F\left (k,T \right)&\xrightarrow[T \rightarrow 0]{}&T S_1(u),\nonumber\\
        F\left (k,T \right)&\xrightarrow[T \rightarrow \infty]{}&T S_2(u),
\end{eqnarray}
at small and large times respectively with

\begin{equation}
    S_1(u)=\frac{2~ e^{-2 u^2} \left(e^{2 u^2}-1\right) }{u^2},
\end{equation}
and
\begin{equation}
     S_2(u)=  \frac{2~ e^{-2 u^2} \left(e^{2 u^2}-e^{g u^2}\right)
   \left(2+ \text{Pe}^2 (1- \bar{\rho})\right)}{(2-g) u^2}. 
\end{equation}
In the asymptotic limits, we thus obtain the following behaviors for the fluctuations,
\begin{eqnarray}
\langle {Q_{\rho}(T)}^2 \rangle_c&\xrightarrow[T \rightarrow 0]{}&\sqrt{T}\frac{\sigma_{\bar{\rho}}}{8\pi}  \int_{-\infty}^{\infty} du \,  S_1(u),
\label{second_cumulantrho_zero_magt0}
\end{eqnarray}
and
\begin{eqnarray}
\langle {Q_{\rho}(T)}^2 \rangle_c&\xrightarrow[T \rightarrow \infty]{}&\sqrt{T}\frac{\sigma_{\bar{\rho}}}{8\pi}  \int_{-\infty}^{\infty} du \,  S_2(u).
\label{second_cumulantrho_zero_magtinfty}
\end{eqnarray}
These integrals can be easily computed yielding the leading behaviors of the variance in the asymptotic limits as in equations~\eqref{rho_cumtinfty_exp}~and~\eqref{rho_cumt0_exp}.

\section{Simulations}
\label{appendix_simulations}
\subsection{Microscopic simulations}

\begin{figure} [!t]
 \includegraphics[width=0.8\linewidth]{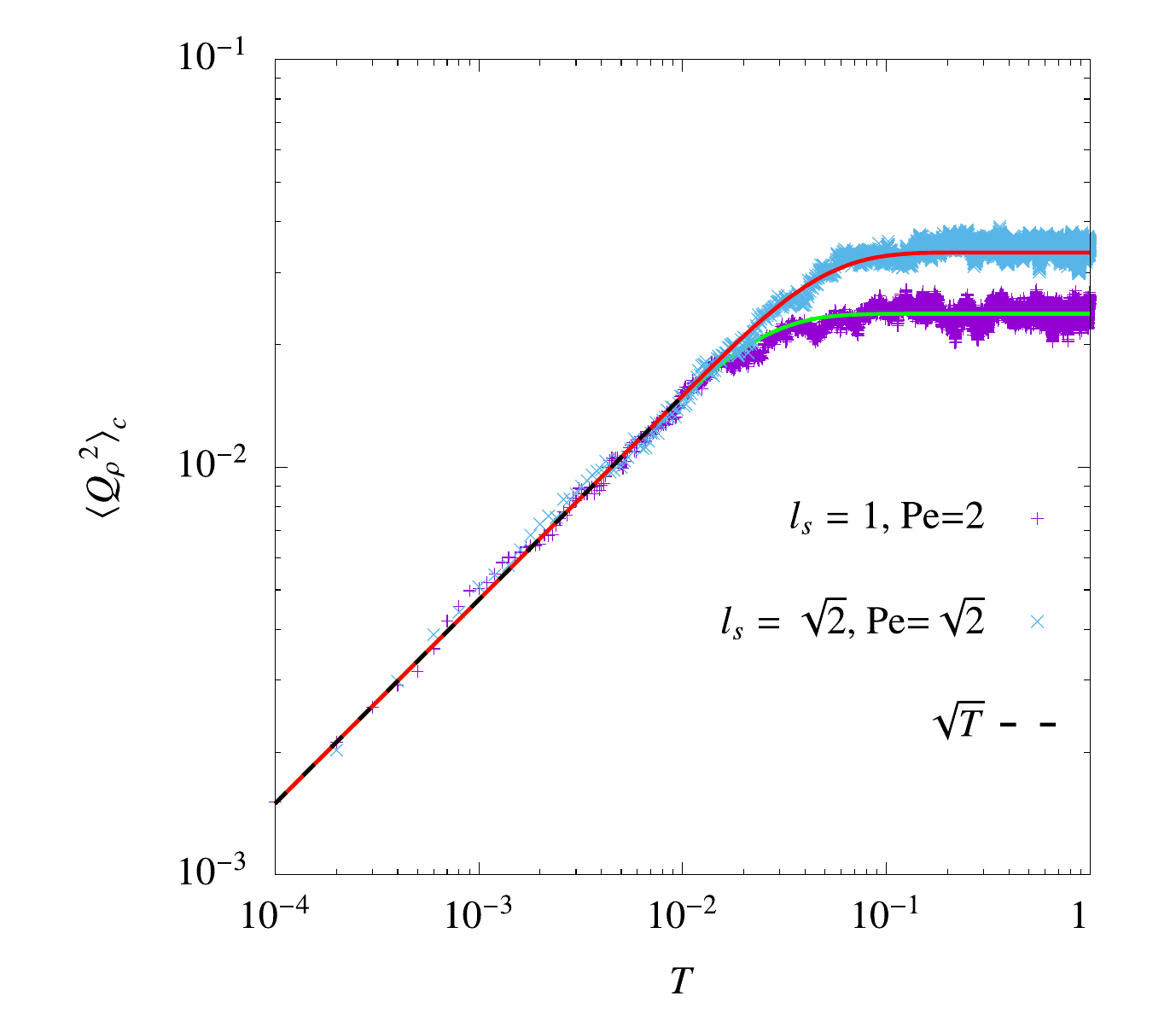}
\caption{
Second cumulant of the integrated density current plotted as a function of time $T$ for finite lattice size $\ell_s$ with periodic boundary conditions. The initial condition is given as $\rho(x,0)=0.25$ and $m(x,t)=0$. The points are obtained from direct Monte Carlo simulations. The solid curves on top of the simulation data correspond to the theoretical result in equation~(\ref{second_cumulantrho_zero_mag}), but with the integral in $k$ replaced by a discrete summation over the Fourier modes ($k_n= 2 n \pi {\ell_s}^{-1}$). We have done the summation numerically over $1000$ modes. Note that the discrete summation measures the current across both boundaries and one has to divide by a factor of $2$ to obtain the actual current across the origin.} The dashed curve corresponds to the small time asymptotic result provided in equation~\eqref{rho_cumt0_exp}. The above plot is in the rescaled coordinates.  
\label{finite_size_effects}
\end{figure}

In the microscopic simulations, we consider a one-dimensional periodic lattice of size $L=1000$ with $N$ particles. The mean density is given as $\bar{\rho}=N/L$. We realize quenched initial profiles by fixing the locations and bias states of the particles at time $t=0$. For the case of a uniform initial density profile, we arrange the particles symmetrically with equally spaced gaps. To obtain zero magnetization initially, the $+$ and $-$ spins (bias states) are also assigned symmetrically. We use the kinetic Monte Carlo method to update the position and states of the particles. To match with the analytical results, we should work in the limit $L \xrightarrow{} \infty, ~\ell_s=\sqrt{\gamma/D} \xrightarrow{} \infty$ with $L/\ell_s$ also $\xrightarrow{} \infty$. However, in microscopic simulations, we use a lattice of finite size and finite values of the rates which introduce deviation from the analytical prediction. These finite lattice effects tend to saturate the cumulant of the integrated current at large times. We can also study finite lattice size effects analytically by treating the Fourier integration in equation~\eqref{second_cumulant_zero_mag} as a discrete sum over finite modes instead of a continuous integration. In figure~\ref{finite_size_effects}, we display the plot of the second cumulant of the integrated density current as a function of time for finite lattice size $\ell_s$ with periodic boundary conditions. We show that the microscopic simulations agree well with equation~\eqref{second_cumulant_zero_mag} with the integration replaced by a discrete sum of the Fourier modes. Additionally, the small-time asymptotics predicted from the infinite lattice analytical calculations can also be matched with the microscopic simulations.
\subsection{Macroscopic simulations}
Since the perturbation equations provided in equations~\eqref{zeroth_order_eqns} and~\eqref{perturbative_eqns} involve non-linear terms, we use the pseudo-spectral method to integrate these equations numerically. Notice that the equations for $\rho_0$ and $m_0$ are solved with initial boundary conditions. These solutions are used to integrate the equations for $p_{\rho_1}$ and $p_{m_1}$ backward in time. Finally, the equations for $\rho_1$ and $m_1$ are integrated forward in time using the solutions for $\rho_0,~m_0,~p_{\rho_1}$~and~$p_{m_1}$. These solutions for the fields at different orders can be used in equation~\eqref{second_cumulant2} to compute the fluctuations for arbitrary boundary conditions. Finite difference schemes also match the microscopic simulations for carefully chosen discretization parameters. In figure~\ref{fig:zeroth_order_fields} of the main text, we have used finite difference schemes with periodic boundary conditions to numerically integrate the hydrodynamic equations with discretization parameters $dx={10}^{-3}$ and $dt={10}^{-8}$. 

\section{References}
\bibliographystyle{unsrt}
\bibliography{bibtex.bib}

\end{document}